\documentclass[useAMS,usenatbib]{mn2e}

\usepackage{graphicx}
\usepackage{amsmath}	
\usepackage{amssymb}	
\usepackage{float}
\usepackage[T1]{fontenc}
\usepackage{ae,aecompl}
\usepackage{color} 
\usepackage[bookmarks=false]{hyperref}
\usepackage[ampersand]{easylist}

\title[Defocusing effects on meteor tracks]{Linear feature detection algorithm for astronomical surveys - II. Defocusing effects on meteor tracks}

\author[D. Bekte\v{s}evi\'{c} et al.]{Dino Bekte\v{s}evi\'{c},$^{1}$\thanks{E-mail: dino@iszd.hr (DB); dejan@iszd.hr (DV); \hbox{arasmus@slac.stanford.edu} (AR)}, Dejan Vinkovi\'{c}$^{1,2}$\footnotemark[1], Andrew Rasmussen$^{3}$\footnotemark[1]  and \v{Z}eljko Ivezi\'{c}$^{4}$\\
$^{1}$Science and Society Synergy Institute, Bana Josipa Jela\v{c}i\'{c}a 22, HR-40000 \v{C}akovec, Croatia\\
$^{2}$Hipersfera d.o.o., Ilica 36, HR-10000 Zagreb, Croatia\\
$^{3}$SLAC National Accelerator Laboratory, 2575 Sand Hill Rd., Menlo Park, CA 94025, USA\\
$^{4}$Department of Astronomy, University of Washington, Box 351580, Seattle, WA 98195-1580, USA}
\begin{document}

\date{Accepted . Received ; in original form }

\pagerange{\pageref{firstpage}--\pageref{lastpage}} \pubyear{2017}

\maketitle

\label{firstpage}

\begin{abstract}
Given the current limited knowledge of meteor plasma micro-physics and its interaction with the surrounding atmosphere and ionosphere, meteors are a highly interesting observational target for high-resolution wide-field astronomical surveys. Such surveys are capable of resolving the physical size of meteor plasma heads, but they produce large volumes of images that need to be automatically inspected for possible existence of long linear features produced by meteors. Here we show how big aperture sky survey telescopes detect meteors as defocused tracks with a central brightness depression. We derive an analytic expression for a defocused point source meteor track and use it to calculate brightness profiles of meteors modeled as uniform brightness disks. We apply our modeling to meteor images as seen by the SDSS and LSST telescopes. The expression is validated by Monte Carlo ray-tracing simulations of photons traveling through the atmosphere and the LSST telescope optics. We show that estimates of the meteor distance and size can be extracted from the measured FWHM and the strength of the central dip in the observed brightness profile. However, this extraction becomes difficult when the defocused meteor track is distorted by the atmospheric seeing or contaminated by a long lasting glowing meteor trail. The FWHM of satellites tracks is distinctly narrower than meteor values, which enables removal of a possible confusion between satellites and meteors.
\end{abstract}


\section{Introduction}

Meteors are atmospheric phenomena caused by hypervelocity impacts of meteoroid particles with planetary atmospheres \citep{Ceplecha}. Meteoroid entry velocities into the Earth's atmosphere range from 11 to 72 km/s \citep{meteorspeed}. Such high levels of kinetic energy transform the compressed air in front of the meteoroid into a hot glowing plasma seen from the ground as a meteor. Under these conditions the meteoroid body is evaporated and ablated. Typically the meteors observed from the ground are caused by sub-centimeter sized particles \citep{risk} at altitudes between about 80 km and 120 km \citep[e.g.][]{altitude}. Larger objects can reach lower altitudes and, if they are of an adequate size and strength, they can partially survive the atmospheric flight in fragments that fall to the ground as meteorites. Such events are sometimes accompanied by ground damages due to a large air burst \citep[e.g. the recent Chelyabinsk event;][]{Chelyabinsk} or cratering from the impacts \citep{Asteroids}. At altitudes above 120 km sputtering can eject particles from the meteoroid surface, which then results in a large air-glow around the flying meteoroid due to thermalisation of fast particles in the atmosphere \citep{sputtering}.

Exploration of meteors provides a valuable insight into the meteoroid origins and their distribution within the Solar System \citep{Asteroids}. Their physical and chemical properties can be connected to their parent bodies and reveal some aspects of the evolution of the Solar System \citep{Borovicka}. Estimates of daily input rate of meteoroids into the Earth's atmosphere range from $\sim$3 to 300 metric tons \citep{chemistry}, which also means a delivery of various non-atmospheric atomic and molecular species high into the atmosphere \citep{Jenniskens}. Hence, interaction of the meteor plasma with the surrounding atmosphere and ionosphere helps in investigations of the physical and chemical properties of the Earth atmosphere and its electrodynamics \citep[e.g.][]{Dyrud,chemistry,Eregion,Oppenheim}.

For a long time meteors have been difficult observational targets for high-resolution high-sensitivity imaging sensors because of their large apparent length on the sky and unpredictable exact time and location of appearance \citep{Bouquet}. The recent proliferation of observational techniques and data volumes, including data collected by amateur astronomers \citep{amateur}, transformed meteor science into a Big Data science field \citep{BigData}. Interestingly enough, while the high-resolution high-sensitivity astronomical sky surveys have revolutionized astronomy with their dramatic increase in the sky coverage and data throughput \citep{Surveys}, they do not have meteor science among their science goals \citep[the upcoming fast transient sky survey using Tome-e Gozen camera is an exception with meteors being included into its initial data processing pipeline;][]{Tomo-e}. Their field of view is large enough to capture a significant number of meteors \citep{wide-field-survey}. This means that existing and upcoming large sky surveys contain or will contain valuable meteor data. Surveys whose CCD pixel's angular size on the sky is less than $\sim$2$\arcsec$ will be able to resolve 1 meter size objects or less at 100 km distance.

The sheer volume of imaging data from such surveys imposes automatic detection of linear features as a necessity. In our previous paper \citep[][ see also the GitHub repository at \url{https://github.com/DinoBektesevic/LFDA}]{LFDA} we described the linear feature detection algorithm fine-tuned for detecting lines in images from large astronomical sky survey databases. This algorithm can be used for extracting meteors from such images, but in that case we have to take into account that meteors appear at distances that make them appear defocused in images from large aperture telescopes focused to infinity. Meteor observations performed by \cite{Subaru} with the Subaru telescope are an example where defocusing was taken into account.
Moreover, linear features can be also caused by artificial satellites, which can contaminate the automatic detection of meteors.

Thus, here we address three important issues in meteor studies utilizing high-resolution high-sensitivity imaging sky surveys:
\begin{easylist}
\ListProperties(Hide=3,Hang=true,Style*=-- )
& the scale of defocusing effect and parameters that affect its strength,
& possible degeneracy in intensity profile shapes between the distance to an object and its physical size,
& and how to distinguish between satellite and meteors solely based on the shape of their linear tracks in images.
\end{easylist}
\noindent
In our analysis we use two survey telescopes as examples: the Sloan Digital Sky Survey (SDSS) telescope with its 2.5 m aperture and the pixel size of 0$\arcsec$.396 on the sky \citep{SDSS} and the upcoming Large Synoptic Survey Telescope (LSST) with 8.4 m aperture and 0$\arcsec$.2 pixel size \citep{LSST}.
In Section \ref{sec:theory} we cover the theory of defocusing that we then apply in Section  \ref{sec:models} on meteors modeled as a point source, a disk of uniform brightness and a complex 3D structure of meteor plasma density. Meteors often leave behind a trail of ionization that glows for some time. In Section \ref{sec:trail} we discuss how this affects the meteor images. Section \ref{sec:satellites} debates how to distinguish satellites from meteors. Section \ref{sec:examples} provides some imaging examples from the SDSS database. In Section \ref{sec:discussion} we argue that meteor distance and physical size can be extracted from the defocused meteor image. We make conclusions in Section \ref{sec:conclusion}.

\begin{figure}
    \includegraphics[width=0.47\textwidth]{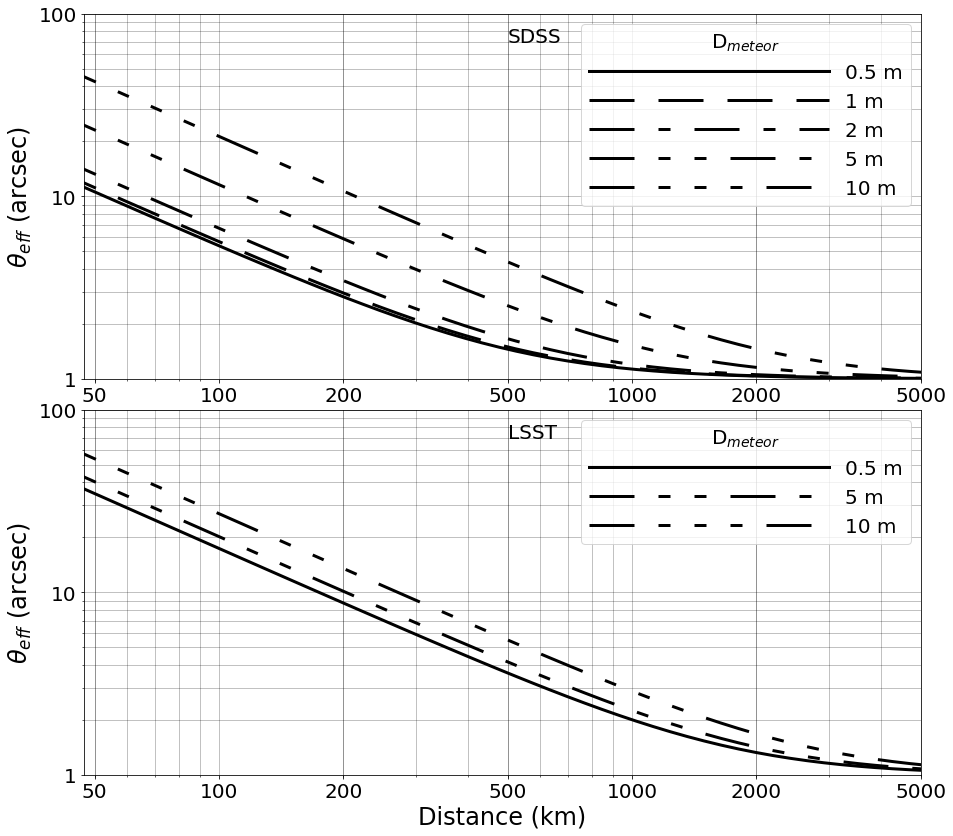}
  \caption{The effective observed FWHM $\theta_{eff}$ is an approximation of the size of meteor track brightness profile (see equation \ref{eq:fullRSS}). The upper graph shows how it changes with the distance to meteors of different size $D_{meteor}$ under 1$\arcsec$ seeing for the SDSS telescope and the lower graph is for the LSST telescope. Notice how $\theta_{eff}$ has the limiting curve when it enters the regime $D_{meteor} \ll D_{mirror}$, which is equivalent to a point source. This approximate approach overestimates the observed FWHM of objects that are bigger than the telescope primary mirror (e.g. see Table \ref{tb:PS_DiffR_SameS_AllH}).}
  \label{fig:approximate}
\end{figure}

\section{Theory of defocusing}
\label{sec:theory}

\subsection{Procedural steps}

Meteors, with their high angular speed, fly through the telescope's field of view instantaneously from the practical point of view (e.g. in SDSS meteors fly over a single camera chip on a time scale of one pixel drift time or less). In broad terms, the meteor light is produced by morphologically two distinct parts \citep[see Fig. B1 in ][]{Hocking}:
\begin{easylist}
\ListProperties(Hide=3,Hang=true,Style*=-- )
& the meteor head (immediate surroundings of the meteoroid body, including the shock front and hot compressed plasma),
& and the meteor wake (the plasma left behind the flying meteoroid that evolves into a glowing trail that diffuses into the surrounding atmosphere).
\end{easylist}
The dominant source of light is the meteor head, which exposes camera pixels as it moves through the field of view. The meteor wake is typically less bright and disappears very quickly, but when it leaves a trail that glows for longer periods of time then it can expose image pixels to values comparable to the meteor head. In our analysis we consider only the meteor head defocus, but in Section \ref{sec:trail} we also discuss deviations from a pure meteor head signal caused by long-lasting meteor trails.

The first step in modeling a detected meteor surface brightness is to decide on the physical model of the meteor head plasma density structure. The physical "size" of a meteor head is actually a poorly defined quantity because it depends on the way how we describe the glowing shape of the meteor head. For example, if we model it as a disk then the size is the disk diameter, but if we use a 3D Gaussian density function then the meteor has no sharp edges and the size can be defined as the function's full width at half maximum (FWHM).

\begin{figure}
  \includegraphics[width=0.5\textwidth]{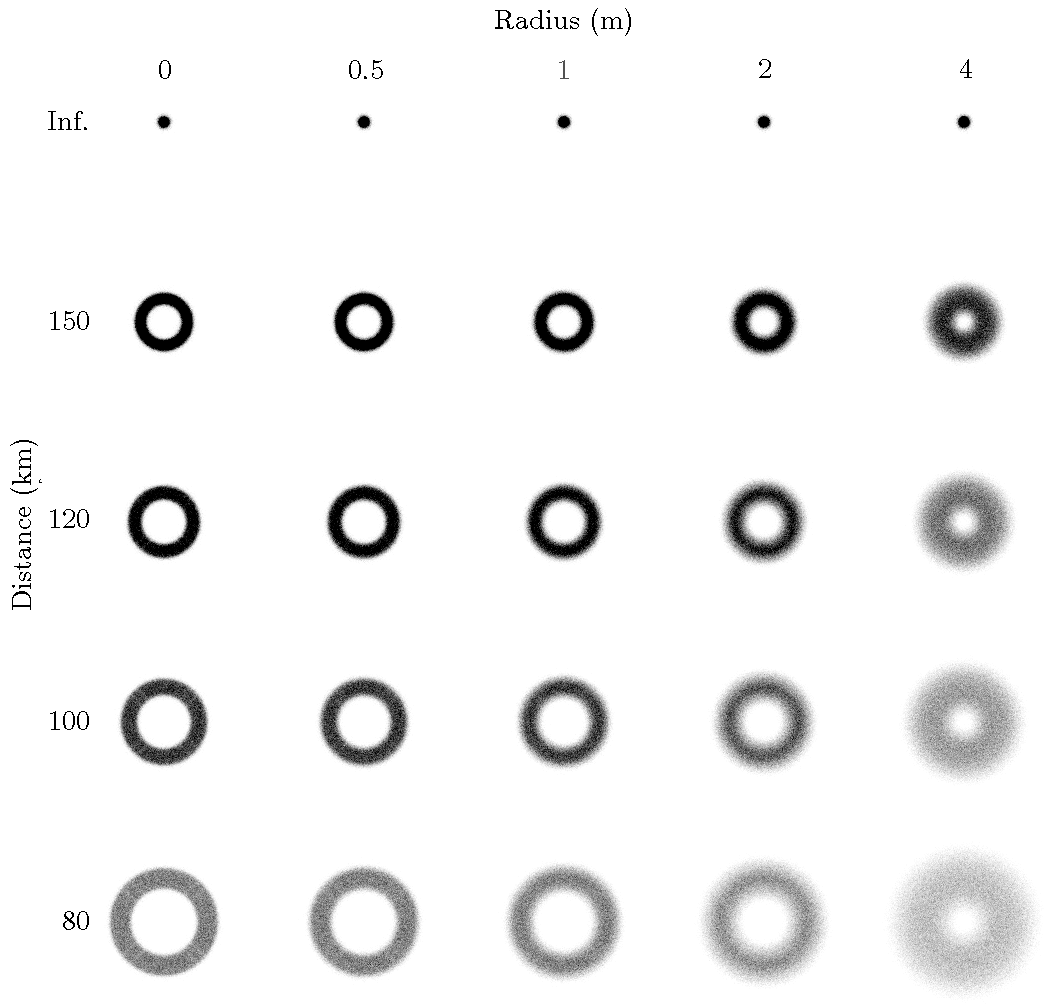}
  \caption{Ray tracing simulations of observed images of static disk sources of various disk radii (horizontal values) at various distances from the telescope (vertical values). Objects in infinity are not defocused and only the atmospheric seeing remains (in this case 0.67$\arcsec$, the top row). Objects at distances typical for meteors appear defocused into ring-like images that increase in size with smaller distances and larger objects. Larger defocus also spreads photons over a larger area such that the surface brightness drops as the defocus effect grows.}
  \label{fig:AltitudeExtentTable}
\end{figure}

\begin{figure}
  \includegraphics[width=0.5\textwidth] {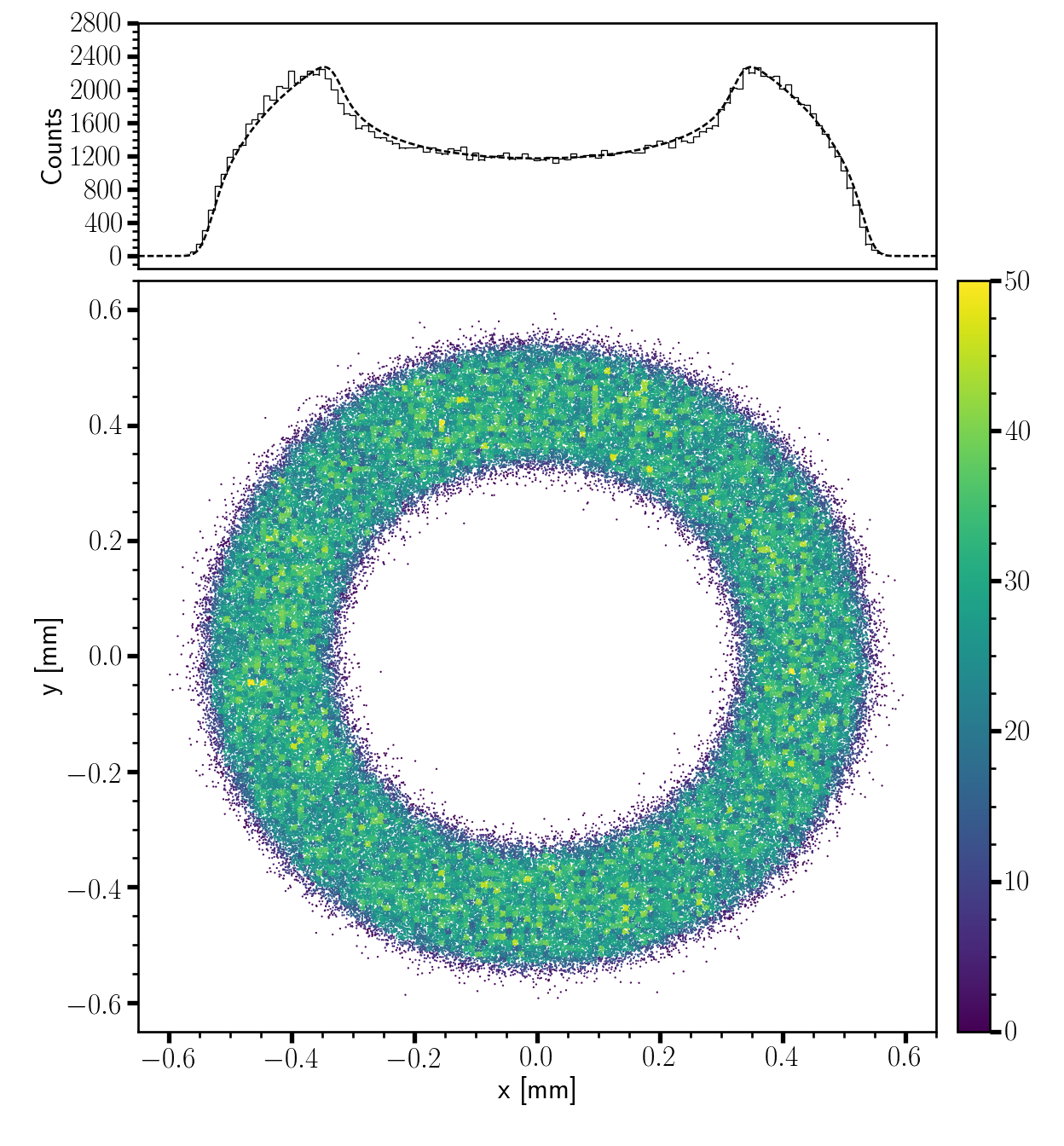}
  \caption{A simulated image (the lower panel) of a static point source at 80~km distance observed with LSST. The light rays produced by the object were convolved with the Kolmogorov seeing of 0.67$\arcsec$ FWHM, while the sky background was set to zero. The rays were traced to the focal plane and binned up on a regular 10~$\micron$ grid to represent LSST pixels. The "donut's" outer diameter is 120 pixels (24$\arcsec$ on the sky). The angular 1D projection (the upper panel) is the brightness profile equivalent to a moving point source at 80~km distance: the thin solid line is a histogram from vertical binning of the image in lower panel, the dashed line is the equivalent theoretical approximation (see equation \ref{eq:pointDefocus} for the analytical function under non-existent seeing).}
  \label{fig:simulation80km}
\end{figure}

\begin{table*}
\caption{FWHM values for a point source at different distances from a telescope under the seeing of $1.48$" for SDSS and $0.67$" for LSST. The defocus FWHM is a result obtained by convolving the meteor by the defocusing function alone (equation \ref{eq:pointDefocus}), while the observed FWHM includes also the Kolmogorov seeing (equation \ref{eq:doubleGaussian}).}
\label{tb:PS_AllH_SameS}
\begin{tabular}{c|c|c|c|c}
&    \multicolumn{2}{c}{SDSS}&    \multicolumn{2}{c}{LSST}\\
Height&    Defocus FWHM&    Observed FWHM&    Defocus FWHM&    Observed FWHM\\
 (km) &        (arcsec)&       (arcsec)&        (arcsec)&       (arcsec)\\
\hline
 80&	5.78&	5.93&	19.80&	19.91\\
100&	4.63&	4.84&	15.84&	15.92\\
120&	3.85&	4.12&	13.20&	13.26\\
150&	3.08&	3.37&	10.56&	10.60\\
\end{tabular}
\end{table*}

This step is followed by applying the atmospheric seeing, i.e. convolving the meteor surface brightness with the atmospheric seeing function. We use the Kolmogorov seeing function approximated with a double-Gaussian profile \citep{Kolmogorov}:
\begin{equation}
p_K (r|\alpha) = 0.909\left[p(r|\alpha) + 0.1p(r|2\alpha)\right]
\label{eq:doubleGaussian}
\end{equation}
where $p(r|\alpha)$ is a single Gaussian given by equation
\begin{equation}
p(r|\alpha) = \frac{1}{2\pi\alpha^2}e^{-\frac{r^2}{2\alpha^2}}
\label{eq:singleGaussian}
\end{equation}
The coefficient $0.909$ comes from the normalization requirement $2\pi\int_0^\infty p_K(r|\alpha)rdr = 1$. The FWHM of Kolmogorov seeing is larger than that of a single Gaussian and it is given by $\text{FWHM} = 2.473\alpha$.

After the light passes through the atmosphere it finally arrives to the telescope where it forms a defocused image influenced by the telescope's aperture. This means that the defocusing convolution is applied after the seeing. In the end, the final meteor image will depend on:
\begin{easylist}
\ListProperties(Hide=4,Hang=true,Style*=-- )
& the inner and outer radius of the primary mirror
& distance to the meteor
& FWHM of the atmospheric seeing
& characteristic size of the meteor
\end{easylist}
\noindent
Meteors leave linear tracks in images and we are interested in brightness profiles across the meteor tracks. It is equivalent to integrating the meteor head image brightness along the observed direction of flight. In our exploration of this parameter space we use FWHM as the measure of brightness profile size under different convolution modalities:
\begin{easylist}
\ListProperties(Hide=3,Hang=true,Style*=-- )
& \textit{Object FWHM} is the meteor angular size without applying the seeing and defocus procedures. Hence, the object profile is the brightness that would produce a meteor observed with a telescope focused on it in an atmosphere without seeing.
& \textit{Defocus FWHM} is the meteor size only due to the defocusing effect, i.e. meteor image under nonexistent seeing.
& \textit{Observed FWHM} is the final result when both the seeing and defocusing are applied.
\end{easylist}

\begin{figure}
  \centering
    \includegraphics[width=0.45\textwidth]{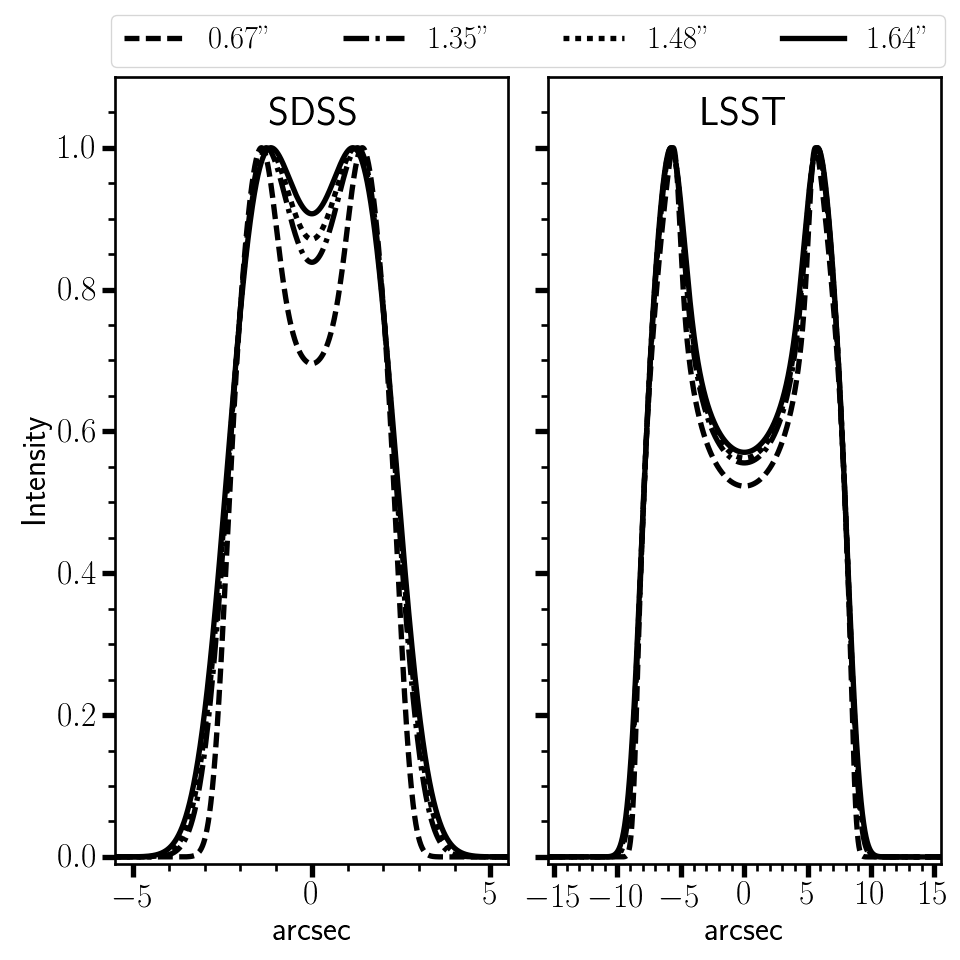}
  \caption{Effects of seeing on the observed intensity profile of a point source located 100km from the imaging instrument. Line types represent results based on different seeing values (as shown in the legend). The profile's inner structure (a dip in the middle) is reduced or completely lost as the seeing worsens. SDSS is much more affected because it has a smaller telescope aperture than LSST.  The overall effect is similar even for much smaller distances to the meteor (see also Fig. \ref{fig:ParameterSpaceSDSS}, \ref{fig:ParameterSpaceLSST} and \ref{fig:ParameterSpaceDEPTH}).}
  \label{fig:Point_seeing}
\end{figure}

\begin{figure}
  \centering
    \includegraphics[width=0.45\textwidth]{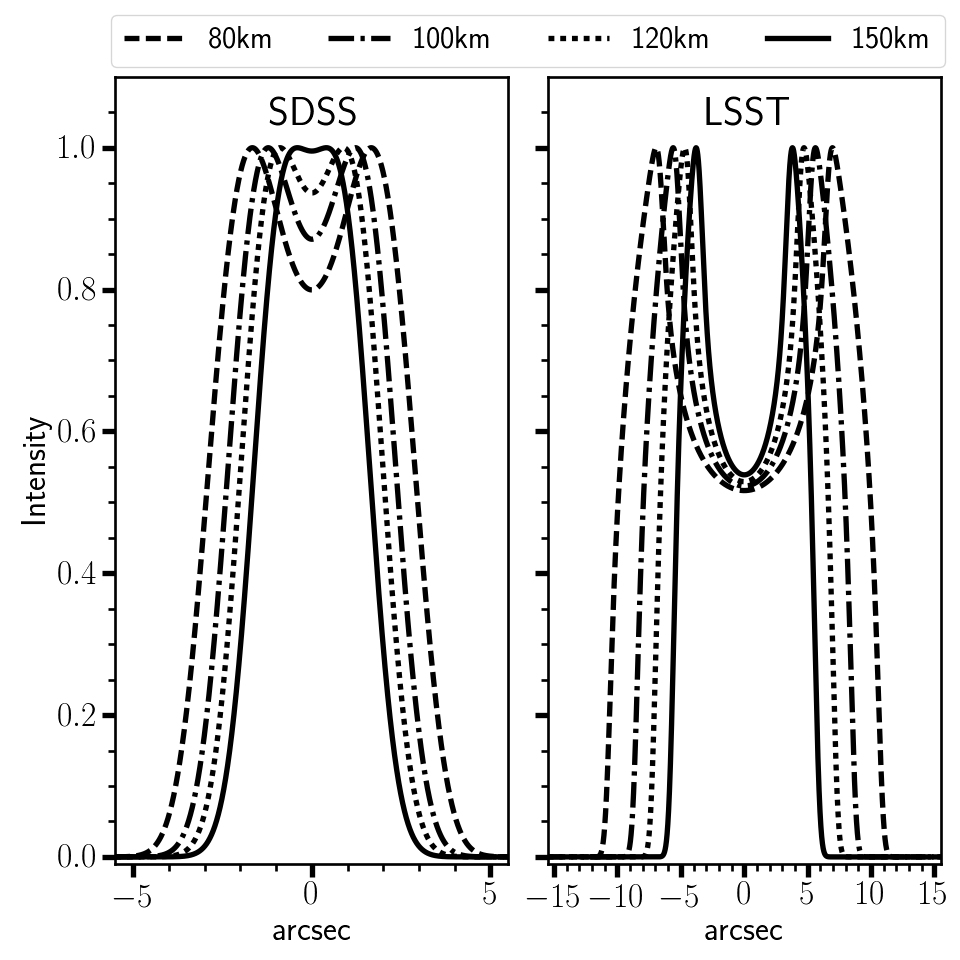}
  \caption{Effects of distance on the observed intensity profile of a point source for a constant seeing ($1.48\arcsec$ for SDSS and $0.67\arcsec$ for LSST).  Line types represent results based on different meteor distances (as shown in the legend). The image shows how smaller distances emphasize the central dip in the profile, until it reaches its extreme value (such as here for LSST) set by the equation \ref{eq.minmax} dictated by the inner and outer radius of the primary mirror (see also Fig. \ref{fig:ParameterSpaceSDSS}, \ref{fig:ParameterSpaceLSST} and \ref{fig:ParameterSpaceDEPTH}).}
  \label{fig:Point_distance}
\end{figure}

\subsection{Approximate solution}

If we ignore for a moment defocusing details in the meteor track profile (discussed further below) and limit our analysis to an effective observed FWHM $\theta_{eff}$ then we can use an approximate equation (in radians)
\begin{equation}
\label{eq:fullRSS}
       \theta_{eff}^2 = \theta_{atm}^2 + {D_{meteor}^2 + D_{mirror}^2 \over d^2},
\end{equation}
where $\theta_{atm}$ is contribution from the seeing and the second term is the defocusing contribution. Here $D_{meteor}$ is the object diameter, $d$ is its distance, and $D_{mirror}$ is the telescope primary mirror diameter. In Fig. \ref{fig:approximate} we show how this effective observed FWHM changes with distance for meteors of different size, from 0.5~m to 10~m, under 1$\arcsec$ seeing. Notice how at satellite distances $\theta_{eff}$ becomes just a few arcsec, close to the seeing alone. We can also see from this approximate analysis that LSST is going to have meteors strongly defocused. Also, in cases when the meteor size is smaller than the mirror size, we can estimate the meteor distance because $\theta_{eff}$ is not very sensitive to the meteor size. This can be understood if we look at two limiting regimes:

\newcounter{qcounter}
\begin{list}{\alph{qcounter})~}{\usecounter{qcounter}
\setlength\labelwidth{0.5cm}
\setlength\leftmargin{0.5cm}}

\item  small meteors: $D_{meteor} \ll D_{mirror}$
\begin{equation}
\label{eq:RSSsmall}
      \theta_{eff}^2 = \theta_{atm}^2 + {D_{mirror}^2 \over d^2},
\end{equation}
Here $D_{meteor}$ is not important; for distant objects, $\theta_{eff} \approx \theta_{atm}$, and for
nearby objects ($d<1000$ km for LSST and $d<500$ km for SDSS) the observed width is essentially the mirror's angular size as seen by the meteor, $\theta_{eff} \approx D_{mirror}/d$.

\item large meteors, $D_{meteor} \gg D_{mirror}$
\begin{equation}
\label{eq:RSSlarge}
      \theta_{eff}^2 = \theta_{atm}^2 + {D_{meteor}^2 \over d^2},
\end{equation}
Here $\theta_{eff} \approx \theta_{atm}$ for large $d$, but for closer objects the observed width is essentially the object's angular size, $\theta_{eff} \approx  D_{meteor} /d$. However, in our detailed modeling of defocused meteor brightness profiles we show that this approximation overestimates the observed FWHM of large objects because of the fat-tail shape of the defocused profile.

\end{list}

\begin{table*}
\caption{Values of FWHM for disk sources of different radii, corresponding to the cases when $R_{meteor}\approx R_{mirror}$  and $R_{meteor}\gg R_{mirror}$. The object FWHM is a measure of the meteor track brightness profile as seen by a telescope focused on the meteor and with no seeing. The defocused FWHM is when a telescope is focused to infinity, but without seeing, while the observed FWHM has the seeing also included (Gauss-Kolmogorov with FWHM of $0.67\arcsec$ for LSST and $1.48\arcsec$ for SDSS). For the case $R_{meteor}\ll R_{mirror}$ see Table \ref{tb:PS_AllH_SameS}. The values follow the rule $\textrm{FWHM}_{\textrm{Object}} < \textrm{FWHM}_{\textrm{Defocus}} < \textrm{FWHM}_{\textrm{Observed}}$, except in the cases when the object is large enough to have the defocusing effects visible only at its edges. In those cases defocusing produces fat-tail effects in the observed brightness profile, which results in $\textrm{FWHM}_{\textrm{Defocus}} < \textrm{FWHM}_{\textrm{Observed}} < \textrm{FWHM}_{\textrm{Object}}$. This is visible in the SDSS case when $R_{meteor}\gg R_{mirror}$ in this table and also in Fig. \ref{fig:DS_3Cases} (the right panel).}
\label{tb:PS_DiffR_SameS_AllH}
\begin{tabular}{c|ccc|ccc}
&    \multicolumn{3}{c}{$R_{meteor}\approx R_{mirror}$}&    \multicolumn{3}{c}{$R_{meteor}\gg R_{mirror}$}\\
Distance&    Object FWHM&    Defocus FWHM&    Observed FWHM&    Object FWHM&    Defocus FWHM&    Observed FWHM\\
 (km) &        (arcsec)&       (arcsec)&       (arcsec)&       (arcsec)&        (arcsec)&       (arcsec)\\
\hline
&    \multicolumn{3}{c}{SDSS ($R_{meteor}$=1~m)}&    \multicolumn{3}{c}{SDSS ($R_{meteor}$=4~m)}\\
\hline
 80&	2.23&	5.97&	6.17&	8.93&	8.44&	8.55\\
100&	1.79&	4.78&	5.00&	7.15&	6.75&	6.89\\
120&	1.49&	3.98&	4.21&	5.95&	5.63&	5.79\\
150&	1.19&	3.19&	3.40&	4.76&	4.50&	4.70\\
\hline
&    \multicolumn{3}{c}{LSST ($R_{meteor}$=4~m)}&    \multicolumn{3}{c}{LSST ($R_{meteor}$=8~m)}\\
\hline
 80&	8.93&	21.77&	21.79&	17.86&	22.15&	22.19\\
100&	7.15&	17.42&	17.43&	14.29&	17.72&	17.76\\
120&	5.95&	14.51&	14.53&	11.91&	14.77&	14.81\\
150&	4.76&	11.61&	11.63&	9.53&	11.81&	11.86\\
\end{tabular}
\end{table*}

\begin{figure}
  \centering
    \includegraphics[width=0.45\textwidth]{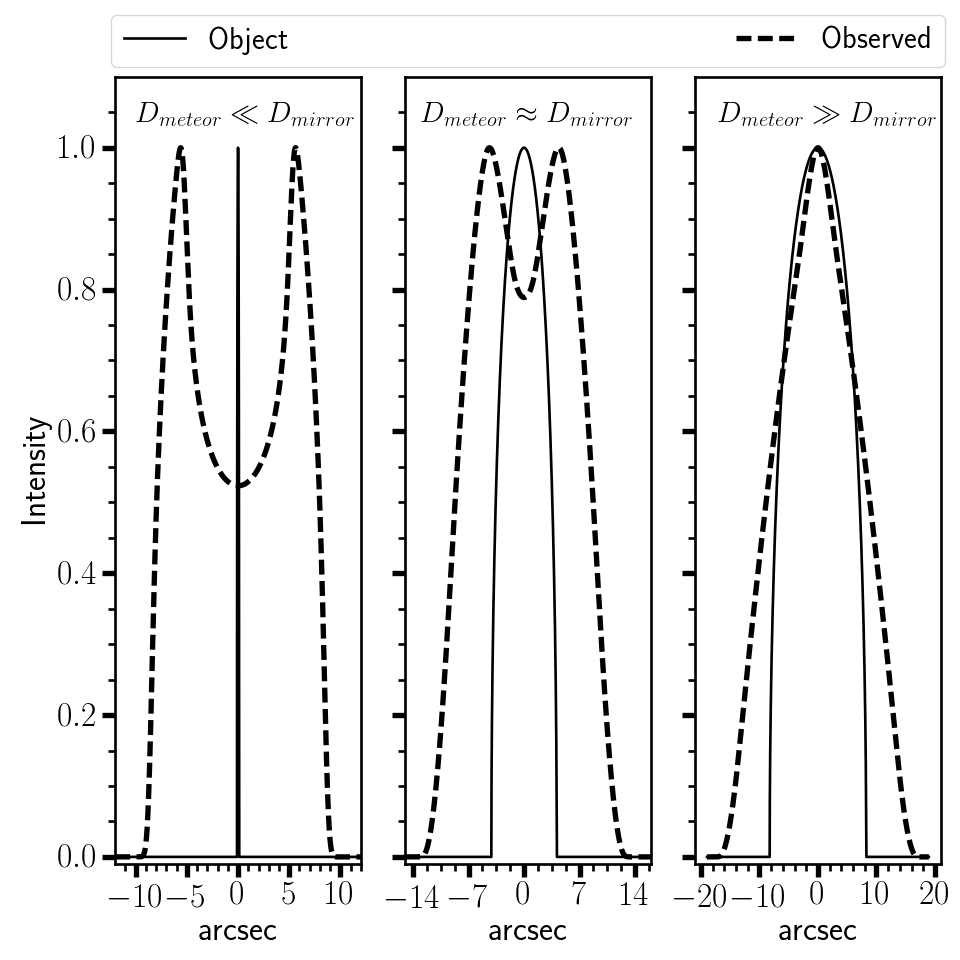}
  \caption{Three cases of meteors with $\theta_D \ll \theta_o$, $\theta_D \approx \theta_o$ and $\theta_D \gg \theta_o$ (see equations \ref{eq:pointDefocus} and \ref{eq:diskDefocus}) illustrated for the LSST telescope at 100~km distance and the seeing of $0.67\arcsec$. Meteors are modeled as disks with a uniform surface brightness and the radii of 0.1~m, 4~m and 8~m, respectively. The solid line shows how the meteor track looks like when the telescope is focused on the meteor without any seeing, while the dashed line shows what we actually see under defocusing and seeing. For a small disk diameter the defocused profile corresponds to that of a point source. As the meteor diameter approaches the inner diameter of the LSST primary mirror, the defocusing profile starts to lose its dip in the middle.}
  \label{fig:DS_3Cases}
\end{figure}

\begin{figure}
  \centering
    \includegraphics[width=0.45\textwidth]{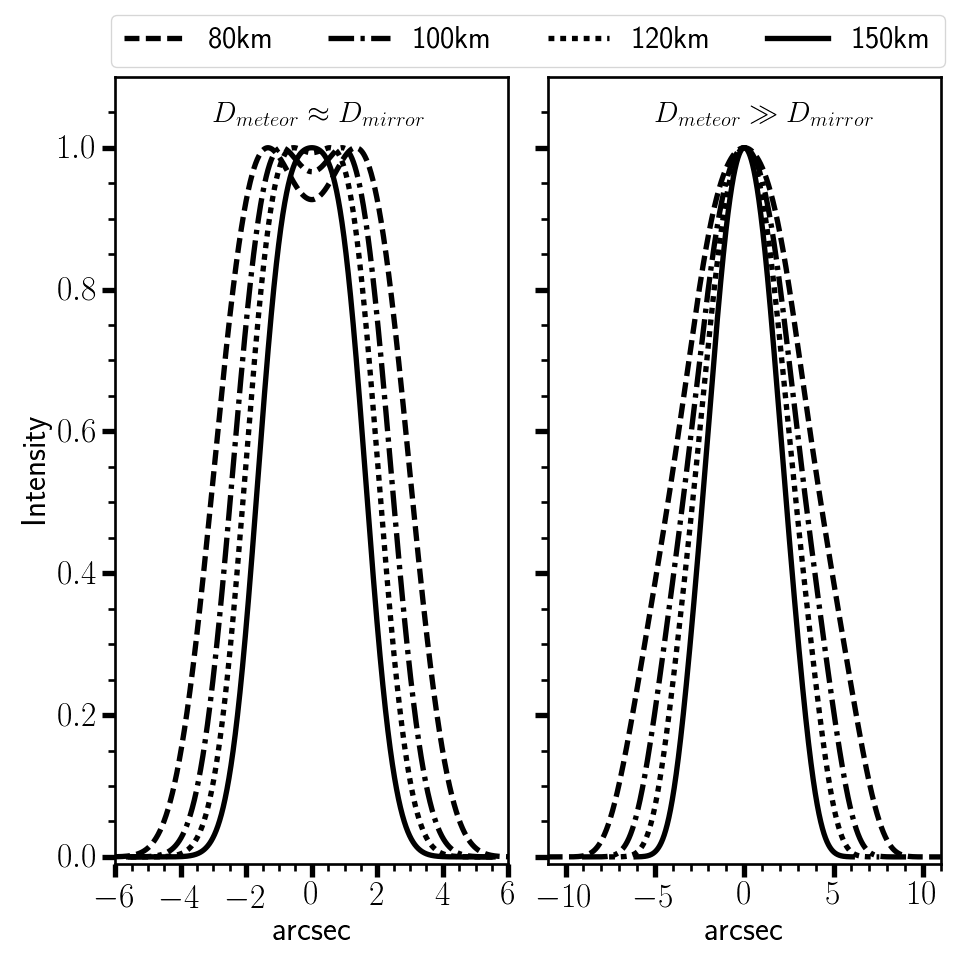}
  \caption{Two cases of uniform brightness disk meteors with $\theta_D \approx \theta_o$ ($R_{meteor}$=1~m) and $\theta_D \gg \theta_o$ ($R_{meteor}$=4~m) illustrated for the SDSS telescope at various distances (different line types) and the seeing of $1.48\arcsec$ (see equations \ref{eq:pointDefocus} and \ref{eq:diskDefocus}). As seen already in Fig. \ref{fig:Point_seeing}, this seeing transforms even a point source into an object similar to $\theta_D$ in size, which results in a defocused image with a negligible central drop in the brightness profile. The distinguishing element for a disk observed with SDSS is the very wide peak when the disk is similar in size to the telescope primary mirror and a growing FWHM as the disk becomes much larger than the mirror. For a small disk diameter ($\theta_D \ll \theta_o$) see Fig. \ref{fig:Point_distance}.}
  \label{fig:Disk_SDSS_objects}
\end{figure}

\begin{figure}
  \centering
    \includegraphics[width=0.45\textwidth]{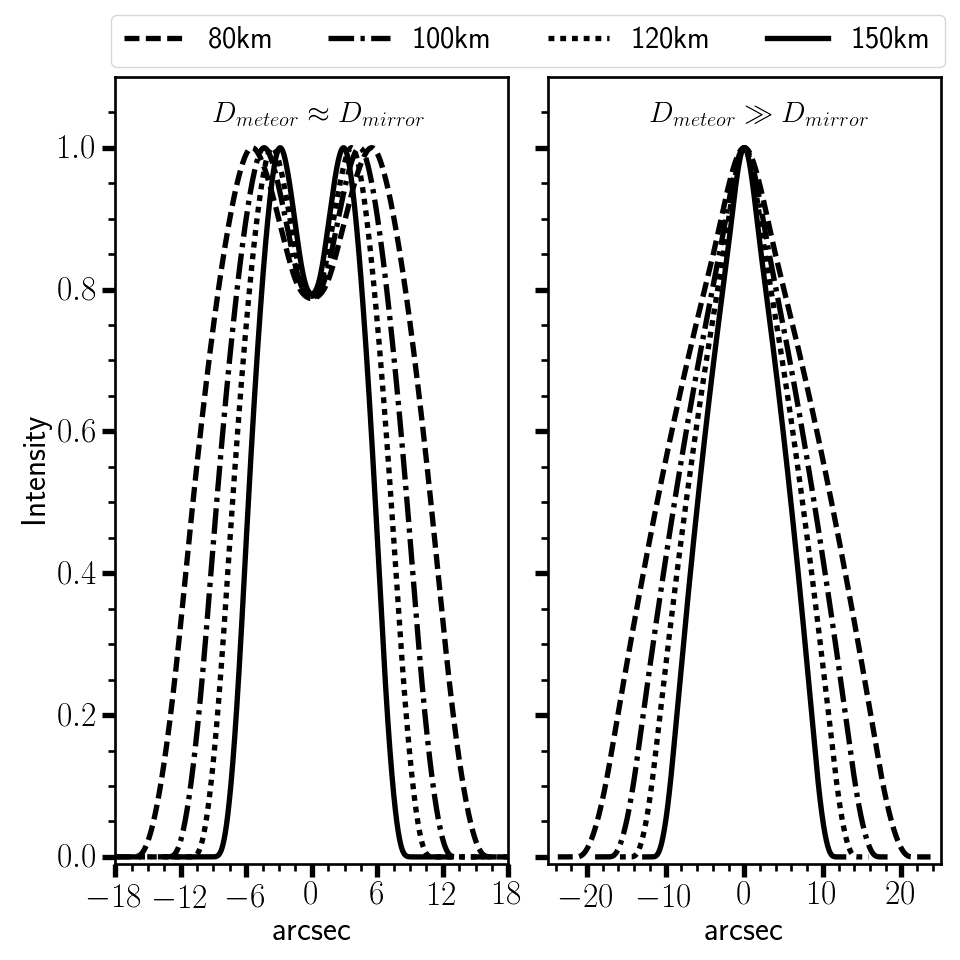}
  \caption{Two cases of uniform brightness disk meteors with $\theta_D \approx \theta_o$ ($R_{meteor}$=4~m) and $\theta_D \gg \theta_o$ ($R_{meteor}$=8~m) illustrated for the LSST telescope at various distances (different line types) and the seeing of $0.67\arcsec$ (see equations \ref{eq:pointDefocus} and \ref{eq:diskDefocus}). Since the seeing FWHM is much smaller than the apparent angular size $\theta_D$ of the disk in the sky, the brightness profiles are dominated by the defocusing effect. For a small disk diameter ($\theta_D \ll \theta_o$) see Fig. \ref{fig:Point_distance}}
  \label{fig:Disk_LSST_distance}
\end{figure}

\begin{figure}
  \centering
    \includegraphics[width=0.45\textwidth]{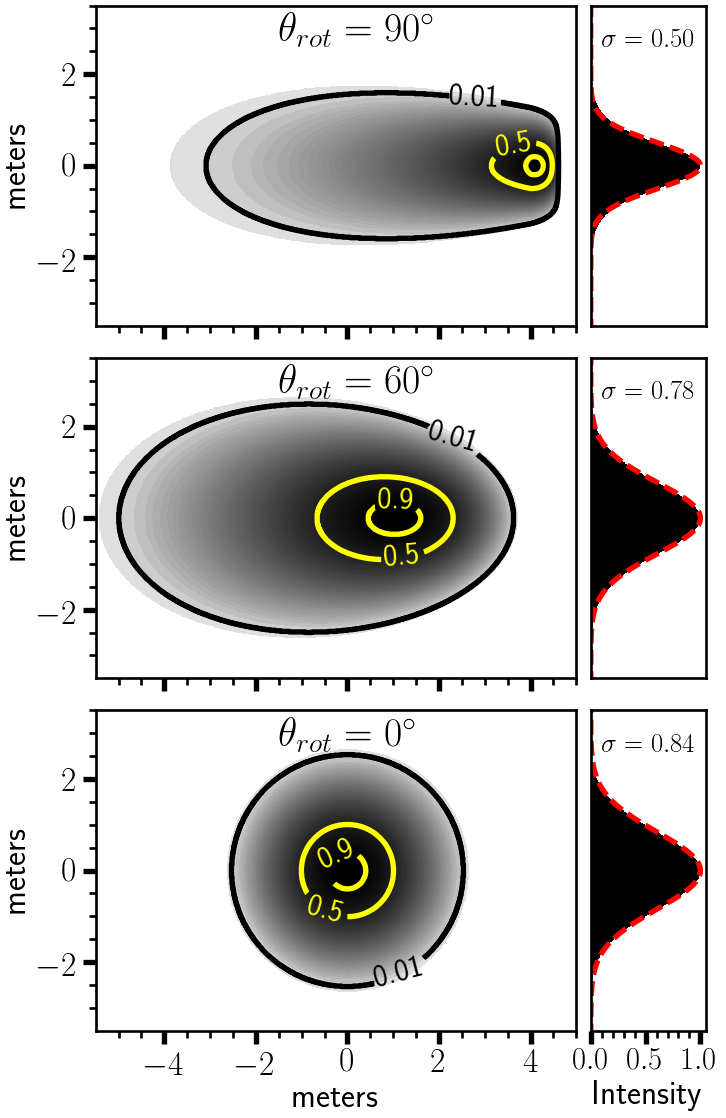}
  \caption{Three examples of a 3D meteor head model (see equation \ref{eq:RabinaDist}) rotated by $90^\circ$, $60^\circ$ and $0^\circ$ relative to the line of sight and projected onto the sky. The intensity values of the images (left panels) are logarithmic in order to highlight the faint features. The contour lines enclose areas of intensity values larger than 1\%, 50\% and 90\% of the maximum intensity value. The 1D brightness profile projections in the direction of meteor flight on the sky are shown in the histograms (right panels). Dashed lines represent fits of the Gaussian function to the 1D brightness profiles with FWHMs of 1.18~m, 1.84~m and 1.98~m (FWHM$\sim$2.355$\sigma$). It shows how differences in the Gaussian widths become tiny for angles smaller than $60^\circ$.}
  \label{fig:Rabina_3Cases}
\end{figure}

\subsection{Analytic treatment}

A defocused image of a finite size object on the sky can be calculated by convolving its brightness profile with the defocus function of a point source. A 2D defocused image of a static point source at distance $d$ is simply a projection of the primary mirror to this distance, converted onto an angular scale. Since a meteor moves through the field of view, the 1D brightness profile across its track on the sky is an integrated defocused 2D brightness in the direction of flight. A point source (i.e. zero-size) meteor is then just a 1D integration of the primary mirror projection. Hence, we can derive a simple analytical expression for a point source meteor as
\begin{equation}
\label{eq:pointDefocus}
    I(\theta) = {2\over \pi \, (\theta_o^2 - \theta_i^2) } \,  \\
     \left[ H(\theta_o-|\theta|) \, \sqrt[]{\theta_o^2 - \theta^2} - H(\theta_i-|\theta|) \,\sqrt[]{\theta_i^2 - \theta^2} \right],
\end{equation}
where $\theta$ is the angular variable measured from the center of the image, $I(\theta)$ is the photon flux per unit solid angle, $H(x)$ is the Heaviside step function ($H(x)=1$ for $x>0$ and 0 otherwise), $\theta_o=R_o/d$, $\theta_i=R_i/d$, $R_o$ is the outer radius of the primary mirror and $R_i$ is the inner radius of the primary mirror (for LSST, $R_o=4180$~mm and $R_i=2558$~mm; for SDSS, $R_o=1250$~mm and $R_i=585$~mm).

For a meteor of finite size, its track's surface brightness profile has to be convolved with the function in equation \ref{eq:pointDefocus}. A realistic modeling has to include an atmospheric seeing, too, which also transforms a point source into a finite size object (convolution of a point with a function yields that same function). Therefore, the complete procedure of reproducing the observed 1D brightness profile cross section of a meteor head track is as follows:
\newcounter{acounter}
\begin{list}{\arabic{acounter})}{\usecounter{acounter}
\setlength\labelwidth{0.35cm}
\setlength\leftmargin{0.35cm}}
\item decide on the 2D surface brightness representation of a meteor (we will use: a point source, a uniform disc, a complicated 2D model dependent on the meteor orientation that essentially yields a 1D Gaussian);
\item integrate the 2D surface brightness along the direction of meteor flight to obtain a 1D brightness profile;
\item apply convolution with the atmospheric seeing function (in our case Kolmogorov seeing in equation \ref{eq:doubleGaussian});
\item convolve this distribution with the defocusing equation \ref{eq:pointDefocus}.
\end{list}

\subsection{Numerical Simulations}

We also performed numerical ray-tracing simulations of meteors observed with LSST. To simulate a point source a ray bundle is created at the entrance aperture of the telescope with aperture position coupled angle according to the distance. Each ray bundle contains approximately $9.2\times10^5$ rays. Rays are evenly distributed in the r-bandpass according to a flat  polychromatic spectral energy distribution. This makes a ray bundle equivalent to an AB 18.5 magnitude source if the integration time is set to 15 seconds. Rays overfill the LSST baffles and primary mirror by approximately 2500~mm such that a spatial distribution could be simulated without causing selection effect artifacts. To simulate spatial extent the rays are repositioned laterally according to a Monte Carlo simulation of the desired surface brightness profile. The seeing is simulated by altering the ray propagation directions prior to their passage through the LSST optical system. Rays optical paths are then simulated. Rays that are not blocked by baffles, spider, etc. have their positions recorded at the plane where the CCD array lies. To produce the final image, positions are binned into 10 microns wide square bins that represent the LSST pixels.

The purpose of simulations was to test how good is our defocusing formula in equation \ref{eq:pointDefocus}. We use uniform disks as meteor shapes and position them at different distances from the telescope. In Fig. \ref{fig:AltitudeExtentTable} we show various examples of simulation results that illustrate how LSST will produce defocused images with distinct differences originating from variations in the meteor size and distance. Since meteors move very fast through the field of view, we integrated these images to compare them with our theory. Fig. \ref{fig:simulation80km} shows an example of such a comparison between a simulated and analytical approach. The example is for a point-source at 80~km distance, where the top panel shows the simulated and analytical 1D meteor track profiles. The lines are almost identical, down to the simulation noise level. This proves the validity of our analytical approach. However, simulations with very high pixel photon levels reveal some second order effects that might be of interest in some future research in cases when the meteors are observed with very high signal-to-noise levels and clearly decoupled from the meteor wake and trail contributions. Examples of such simulations are shown in Appendix \ref{AppendixA} and \ref{AppendixB}.

\begin{table*}
\caption{Values of FWHM for the fiducial 3D meteor model rotated by $90^\circ$, $60^\circ$ and $0^\circ$ as shown in Fig. \ref{fig:Rabina_3Cases}. See Table \ref{tb:PS_DiffR_SameS_AllH} for the column description.}
\label{tb:RS_SameR_SameS_AllH}
\begin{tabular}{c|c|c|c|c|c}
&&    \multicolumn{2}{c}{SDSS}&    \multicolumn{2}{c}{LSST}\\
Distance&    Object FWHM&    Defocus FWHM&    Observed FWHM&   Defocus FWHM&    Observed FWHM\\
 (km) &        (arcsec)&       (arcsec)&       (arcsec)&        (arcsec)&       (arcsec)\\
\hline
&&  \multicolumn{2}{c}{$\theta_{rot}=90^\circ$} &&\\
\hline
 80&	2.55&	6.31&	6.39&	20.02&	20.05\\
100&	2.04&	5.05&	5.13&	16.02&	16.05\\
120&	1.70&	4.21&	4.30&	13.35&	13.38\\
150&	1.36&	3.37&	3.49&	10.68&	10.72\\
\hline
&& \multicolumn{2}{c}{$\theta_{rot}=60^\circ$} &&\\
\hline
 80&	3.94&	6.59&	6.69&	20.43&	20.46\\
100&	3.16&	5.27&	5.40&	16.35&	16.38\\
120&	2.63&	4.39&	4.54&	13.62&	13.66\\
150&	2.10&	3.51&	3.71&	10.90&	10.94\\
\hline
&& \multicolumn{2}{c}{$\theta_{rot}=0^\circ$} &&\\
\hline
 80&	4.25&	6.63&	6.74&	20.50&	20.54\\
100&	3.40&	5.31&	5.44&	16.40&	16.44\\
120&	2.84&	4.42&	4.59&	13.67&	13.71\\
150&	2.27&	3.54&	3.75&	10.93&	10.98
\end{tabular}
\end{table*}

\begin{figure}
  \centering
    \includegraphics[width=0.45\textwidth]{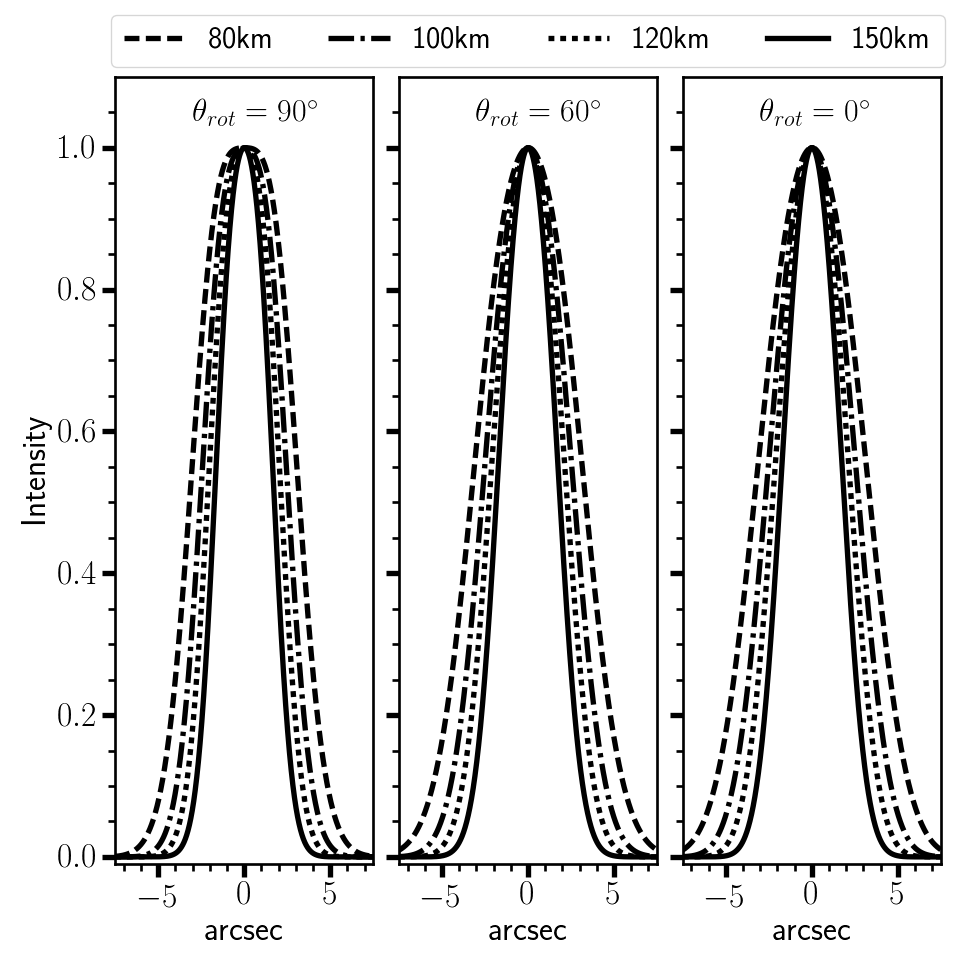}
  \caption{Three cases of the fiducial 3D meteor model rotated by $90^\circ$, $60^\circ$ and $0^\circ$ as shown in Fig. \ref{fig:Rabina_3Cases}, observed with the SDSS telescope from different distances (line types as shown in the legend) under the seeing FWHM of 1.48".}
  \label{fig:RabinaSDSS}
\end{figure}

\begin{figure}
  \centering
    \includegraphics[width=0.45\textwidth]{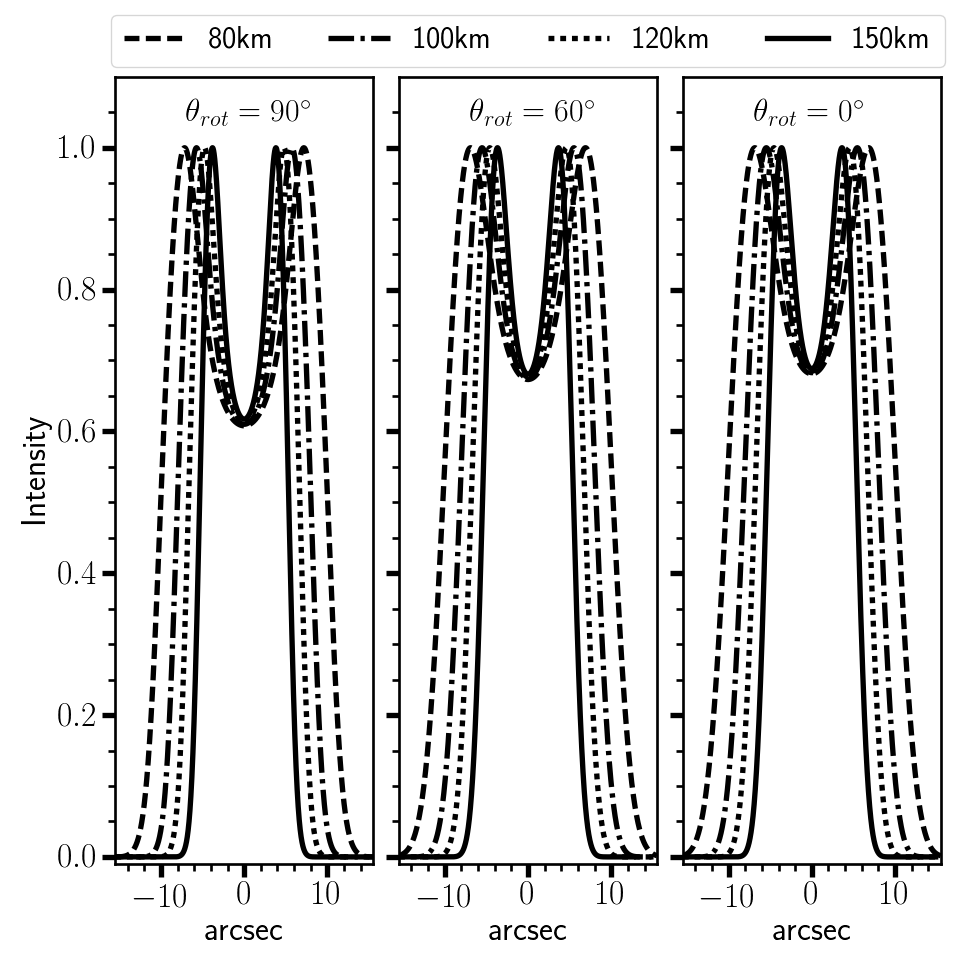}
  \caption{Three cases of the fiducial 3D meteor model rotated by $90^\circ$, $60^\circ$ and $0^\circ$ as shown in Fig. \ref{fig:Rabina_3Cases}, observed with the LSST telescope from different distances (line types as shown in the legend) under the seeing FWHM of 0.67".}
  \label{fig:RabinaLSST}
\end{figure}

\section{Defocusing of different meteor head models}
\label{sec:models}

\subsection{Point Source}

In the regime of small meteors ($D_{meteor} \ll D_{mirror}$) we can use a point source, i.e. zero-size, as the meteor brightness model. We showed above that the approximate equation \ref{eq:fullRSS} yields the observed FWHM to be $\approx D_{mirror}/d$, but the analytic expression in equation \ref{eq:pointDefocus} suggests that the observed profile can deviate significantly from a Gaussian profile. The shape of observed FWHM should critically depend on the amount of seeing and the meteor distance. In Fig. \ref{fig:Point_seeing} we show analytic calculations of the impact that different seeing values have on the observed profile for a given distance of 100~km.  For LSST we use the seeing of 0.67$\arcsec$ representing the currently measured median seeing at the Cerro Pach\'{o}n site of the future LSST observatory \citep{LSST}. For SDSS we use the seeing of $1.35\arcsec$, $1.48\arcsec$ and $1.64\arcsec$ representing the measured\footnote{\url{http://www.sdss.org/dr13/imaging/other_info/}} upper quartile, median and lower quartile values of the SDSS seeing\footnote{The measured seeing is defined as the effective width of the point spread function, where width=1.035$\times$FWHM for Kolmogorov seeing}, respectively. The figure shows that seeing reduces the central drop in the defocused brightness profile. The effect is much stronger for SDSS than LSST because of the SDSS's smaller aperture size. However, the relative differences of seeing effects are small for the realistic ranges of seeing used in Fig. \ref{fig:Point_seeing}. Hence, for practical reasons all the graphs below are made with the appropriate median seeing ($0.67\arcsec$ for LSST and $1.48\arcsec$ for SDSS).

A much stronger impact to the brightness profile comes from variations of the object's distance to the imaging instrument. Fig. \ref{fig:Point_distance} shows how distance to point sources affects the observed defocused brightness profile. Closer point sources have a more pronounced central depression, which reaches its deepest value set by the local maximum and minimum of equation \ref{eq:pointDefocus} under no-seeing conditions:
\begin{equation}
 \label{eq.minmax}
 \frac{I(center)}{max(I)}=\sqrt[]{\frac{1-\epsilon}{1+\epsilon}},
\end{equation}
where $\epsilon=R_i/R_o$. This deepest value is 0.60 for SDSS and 0.49 for LSST. More importantly, the observed FWHM scales with the distance, as expected from the approximate solution in equation \ref{eq:RSSsmall}. For practical purposes we also list the values of defocused FWHM (no seeing) and observed FWHM (with seeing applied) in Table \ref{tb:PS_AllH_SameS}.

\begin{figure}
  \centering
    \includegraphics[width=0.45\textwidth]{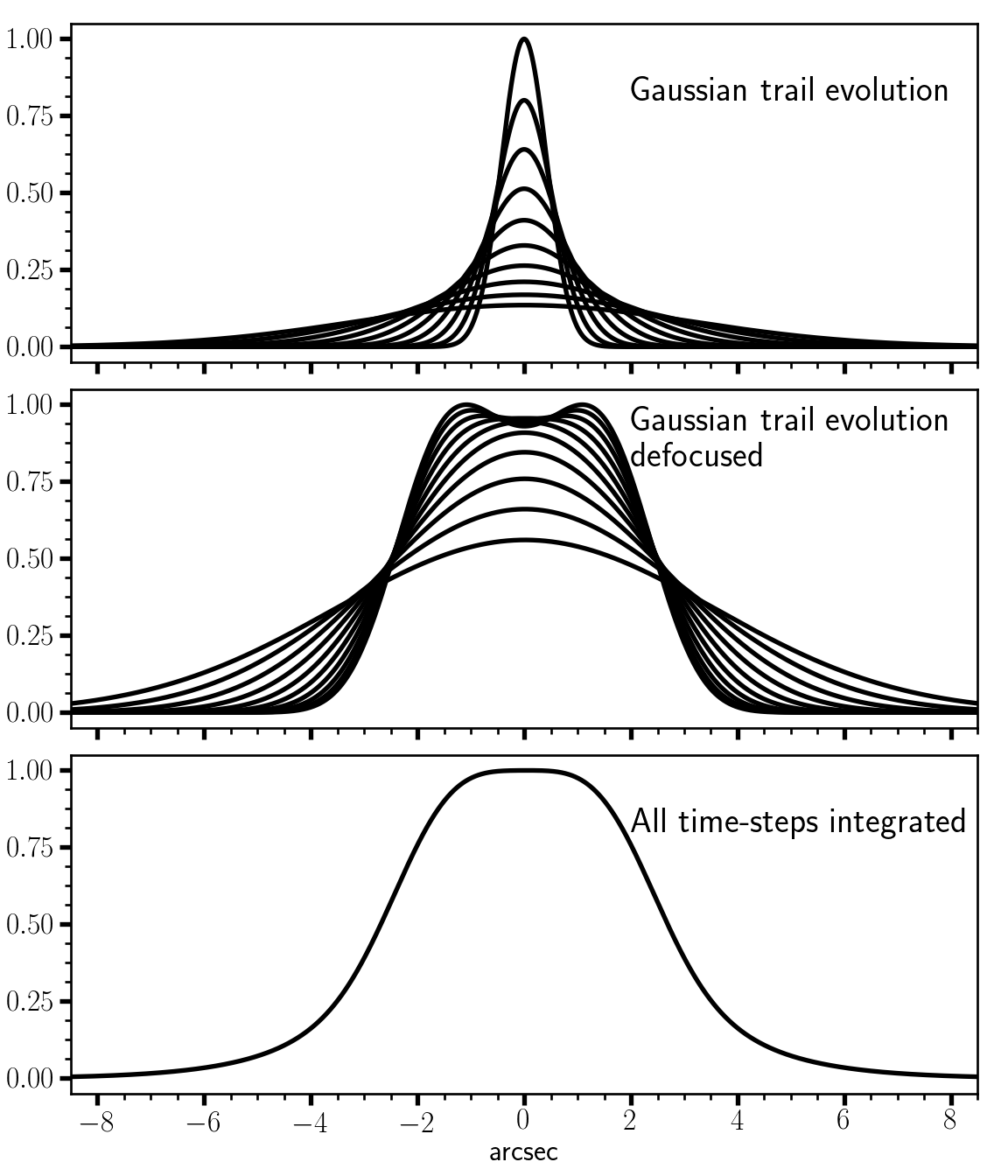}
  \caption{A fiducial model of ionized meteor trail evolution as seen by SDSS at 100~km distance. The top panel is the trail brightness as seen by a telescope focused on the trail without seeing. Different lines show the trail evolution every 0.22 seconds, with the peak brightness evolving as $e^{-t/\tau}$, with $\tau$=1~s, starting from $t$=0~s, while the total emitted light remains the same (i.e. the surface below the curves remains constant). The middle panel shows those profiles convolved with the seeing of $1.48\arcsec$ and defocusing. The bottom panel is the total time integrated trail brightness profile that we actually measure in images. This final curve should be added to the meteor head brightness profile in order to reconstruct the overall meteor track seen in the image. All panels have the maximum brightness scaled to one for clarity. }
  \label{fig:meteor_trailSDSS}
\end{figure}

\begin{figure}
  \centering
    \includegraphics[width=0.45\textwidth]{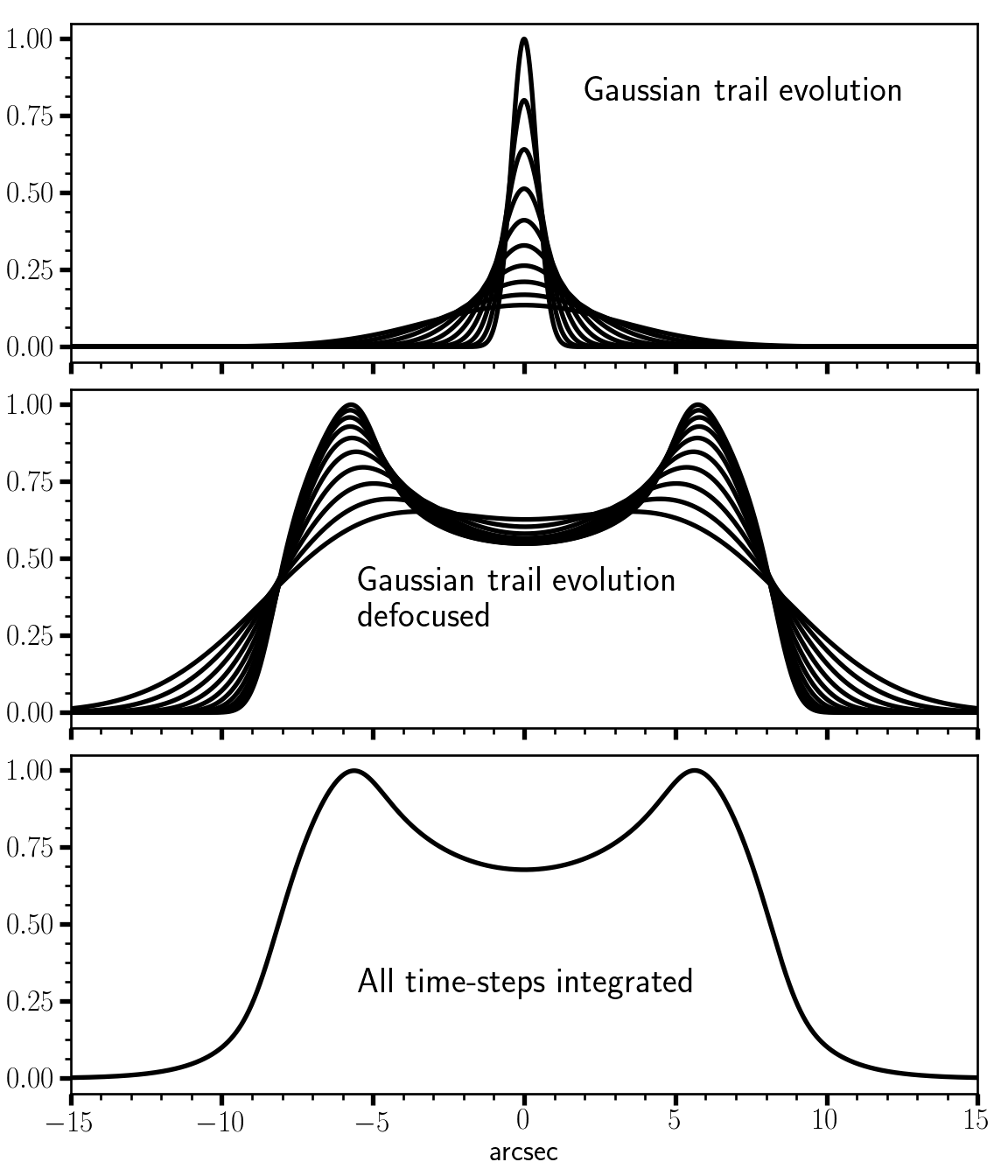}
  \caption{The same as in Fig.\ref{fig:meteor_trailSDSS}, but for the LSST telescope and with the seeing of  $0.67\arcsec$. Here the defocus effect is much stronger than in SDSS due to a larger telescope aperture and now even meteor trails can have a central dip in the brightness profile.}
  \label{fig:meteor_trailLSST}
\end{figure}

\subsection{Disk Model}
\label{sec:disk_model}

Since meteors actually have a finite size, a more realistic approach to the exploration of meteor tracks in images is to attach some surface brightness model to the meteor head. Here we use the simplest approach of a uniform brightness disk. The size of such a disk is then treated as the size of meteor head. The light  profile across the meteor path for such a uniform disk with angular radius $\theta_D=R_D/d$, where $R_D$ is the disk radius at distance $d$, is given by
\begin{equation}
\label{eq:diskDefocus}
   I_D(\theta) \propto \, (\theta_D^2 - \theta^2)^{1/2},
\end{equation}
where $\theta$ is the angular distance from the middle of the track in the sky, and $|\theta| \le \theta_D$. Fig. This profile is then convolved with the seeing and defocusing functions. \ref{fig:DS_3Cases} shows how the disk size influences the observed brightness profile in three possible regimes ($\theta_D \ll \theta_o$, $\theta_D \approx \theta_o$ and $\theta_D \gg \theta_o$) in the case of LSST for a meteor at 100~km distance. For small meteors the convolved profile will differ little from the point source profile, but in meteors larger than the telescope's primary mirror the meteor size dictates the observed FWHM. The observed profile of such large meteors displays a strong fat-tail shape effect, which can lead to a situation where the observed FWHM is slightly smaller than the object FWHM. This is further illustrated in Figs. \ref{fig:Disk_SDSS_objects} and \ref{fig:Disk_LSST_distance} in which the observed profiles for fixed disk sizes and varying distance to the meteor are displayed. In Table \ref{tb:PS_DiffR_SameS_AllH} we list the values of object, defocus and observed FWHM for illustrative examples used in the figures.

\subsection{Fiducial 3D model}
\label{sec:gaussian3D}

Meteor head emission has a far more complicated structure than a simple uniform disk brightness, but the details are still a matter of active research. Hence, here we show defocusing calculations for an analytical 3D model \citep{Rabina} that contains typical meteor head features like a shock front, high density stagnation zone behind the shock, and the density drop immediately behind the meteor. The original equation describes the plasma frequency distribution, but for our needs we adopted the following expression as a 3D distribution of light production
\begin{equation}
\label{eq:RabinaDist}
   I(x, y, z) \propto F(x) \frac{r^2}{\left( r+k(x_0-x) \right)^2}e^\frac{y^2+z^2}{\left( r+k(x_0-x) \right)^2} \\
\end{equation}
\noindent where $F(x)$ is given by:
\begin{align}
\label{eq:RabinaDistF}
 F(x)=
\begin{cases}
e^\frac{x-x_0}{l}, & x\in \langle -\infty, x_0 \rangle\\
1-\frac{(x-x_0)^2}{r^2},  & x\in [x_0, x_0+r]\\
0, & x\in \langle  x_0+r, \infty\rangle
\end{cases}
\end{align}

We use the same parameter values as in the original paper by \cite{Rabina}: $r$=0.6~m, $k$=0.15~m, $l$=3~m, $x_0$=4~m. Our modeling is applied on three cases of the viewing angle relative to the line of sight, $0^\circ$, $30^\circ$ and $90^\circ$, as shown in Fig. \ref{fig:Rabina_3Cases} (left panels). When projected along the direction of meteor flight on the sky, all three cases yield a Gaussian profile (shown in the right panels in Fig. \ref{fig:Rabina_3Cases}). These projections are then convolved with the seeing and defocusing for various distances from the telescope. The results are shown in Figs. \ref{fig:RabinaSDSS} and \ref{fig:RabinaLSST} and their FWHMs are listed in Table \ref{tb:RS_SameR_SameS_AllH}. In SDSS examples there is no visible central dip in the brightness profile and the peak is rather sharp. The LSST examples show the typical central dip that becomes smaller as the Gaussian of projected meteor increases. The viewing angle makes a noticeable difference only when the angle is larger than about 60$^\circ$ and even then only in the cases where the meteor was small enough compared to the primary mirror to produce the central brightness dip associated with the defocusing effects. In general, we see that brightness profiles originating from uniform disk models and Gaussian models are very similar, which makes their distinction difficult.

\begin{figure}
  \centering
    \includegraphics[width=0.45\textwidth]{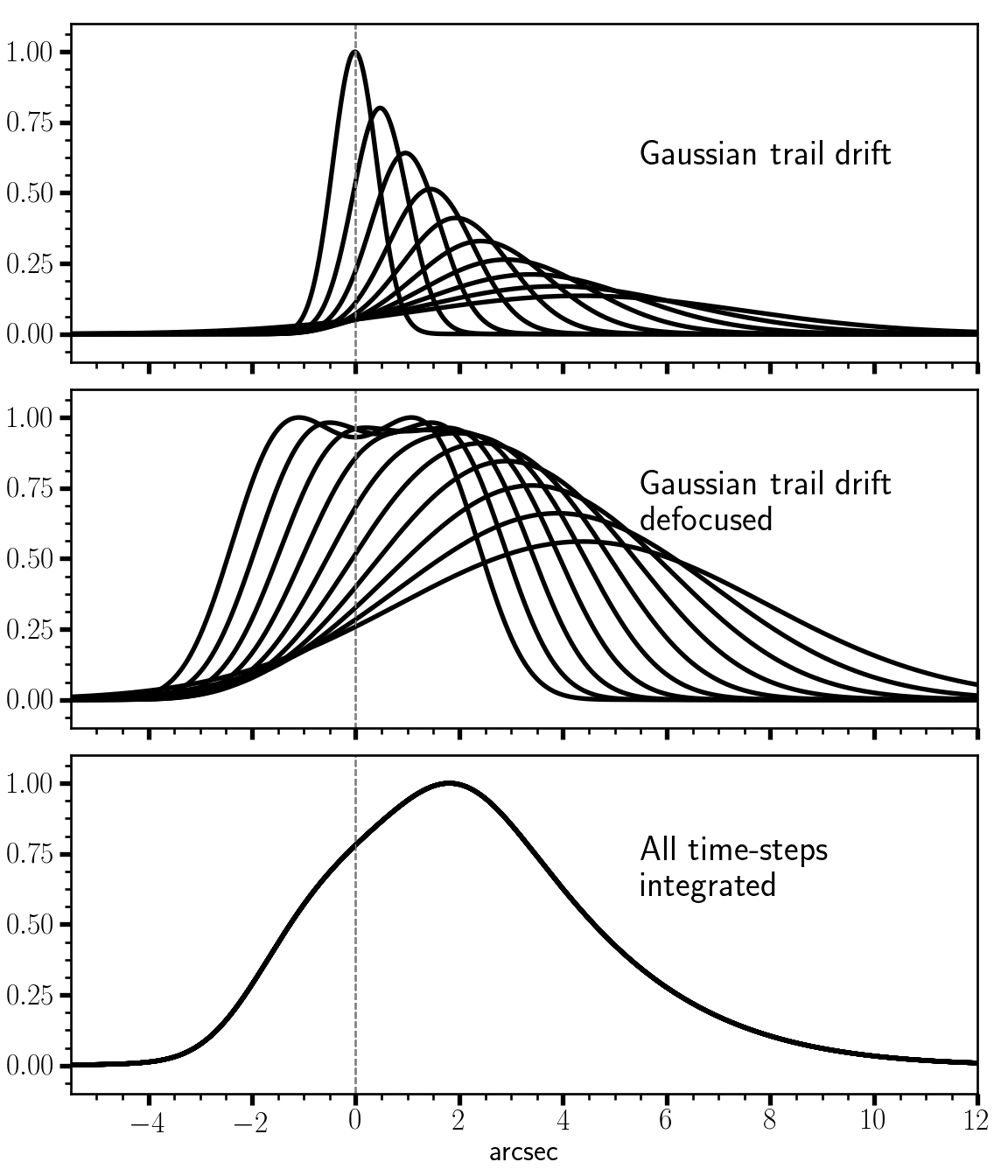}
  \caption{The same as in Fig.\ref{fig:meteor_trailSDSS}, but for a trail that drifts due to atmospheric winds. The vertical dashed line shows the initial position of the meteor trail. The drift speed in this example is $0.486\arcsec$ in each time step (i.e. $2.187\arcsec/s$). The final image profile of such a trail (the bottom panel) shows brightness asymmetries, which results in an asymmetric double-peaked meteor track when combined with a meteor head brightness (see Fig. \ref{fig:addedMeteorTrailSDSS}).}
  \label{fig:meteor_traildriftSDSS}
\end{figure}

\begin{figure}
  \centering
    \includegraphics[width=0.45\textwidth]{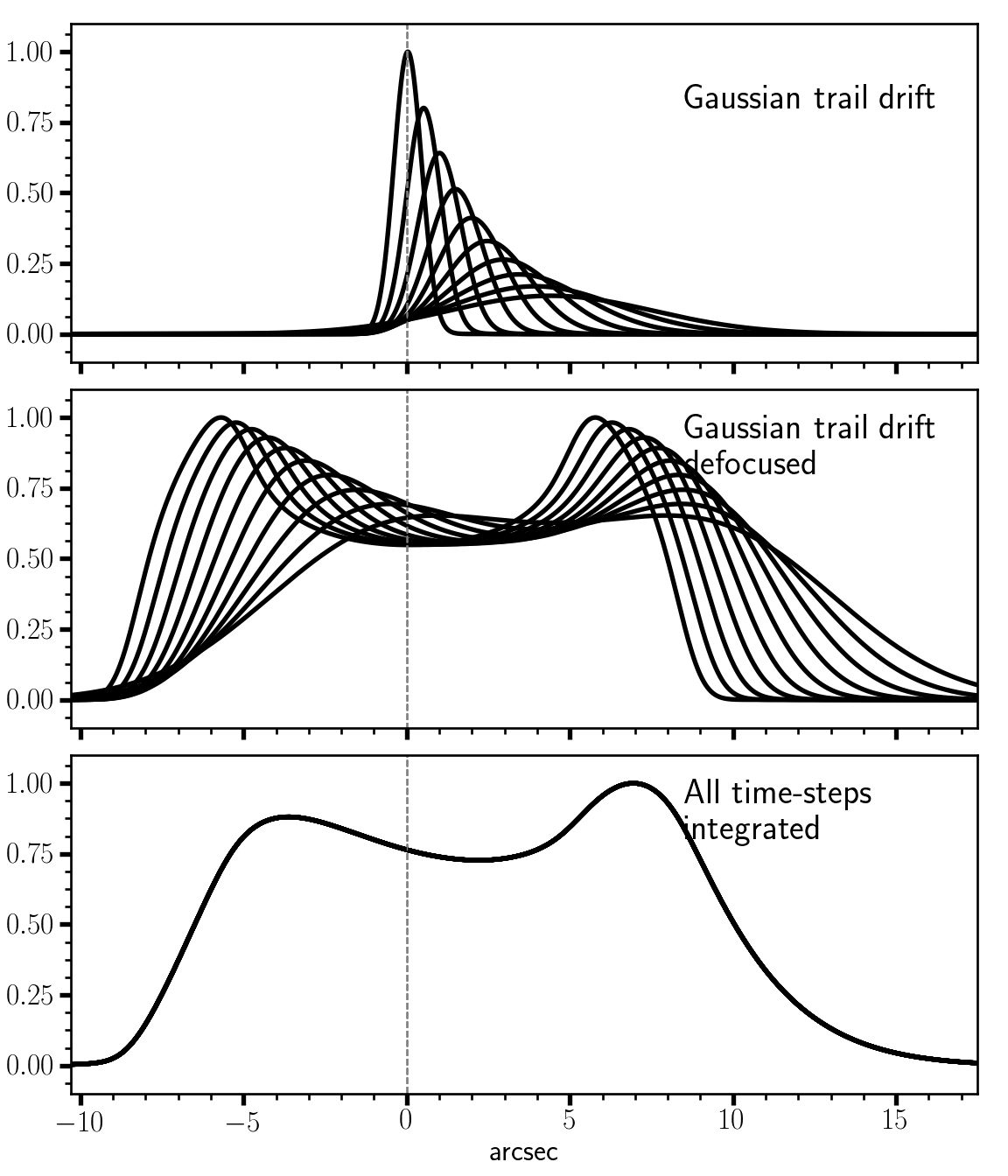}
  \caption{The same as in Fig.\ref{fig:meteor_traildriftSDSS}, but for the LSST telescope. Combination of this trail with a meteor head profile is shown in Fig. \ref{fig:addedMeteorTrailLSST}.}
  \label{fig:meteor_traildriftLSST}
\end{figure}

\subsection{The role of sky background and fragmentation}

In all our calculations we use zero sky background level and perfect instrument sensitivity, which enables reconstruction of pure meteor track brightness profiles. In reality light detectors have some sensitivity cutoff and the sky has some brightness, too. These two effects produce a certain level of sky background in images, which affects primarily the low brightness wings of meteor profiles, but it can also alter the entire shape of low brightness meteor profiles. Hence, extracting meteor data from fits to the measured meteor tracks in astronomical images has to be done with a careful consideration of deviations from the prefect brightness profiles.

This process will affect not only the measured FWHM, but also the depth of the central brightness dip. Our analysis (see equation \ref{eq.minmax}) shows that meteors in SDSS images can not show dips larger than $\sim$40\% of the peak brightness for very close meteors, and $\lesssim$20\% for more realistic distances. If the sky background is very high relative to the meteor peak brightness (i.e. the meteor brightness is very low, as it can be for very small, point source meteors) then the dip will appear reduced compared to the analytical expectation.

However, a large central brightness dip can be an indication of a meteor fragmentation, where two pieces are flying in close vicinity in parallel. In such a case three additional effects will be present: a) the overall FWHM must be quite large, because brightness profiles of the two pieces partially overlap and produce an impression that they are one big meteor, b) the central dip becomes enhanced beyond the limits presented in our analysis because the origin of this brightness depression is not only defocusing, but also a physical separation between the meteor pieces, and c) each side of the double brightness peak can have its own small dips because tiny meteor pieces are defocused into their individual profiles with the central dips. We will illustrate this effect of fragmentation in Section \ref{sec:examples} below using an example of meteor in SDSS images.

\begin{figure}
  \centering
    \includegraphics[width=0.45\textwidth]{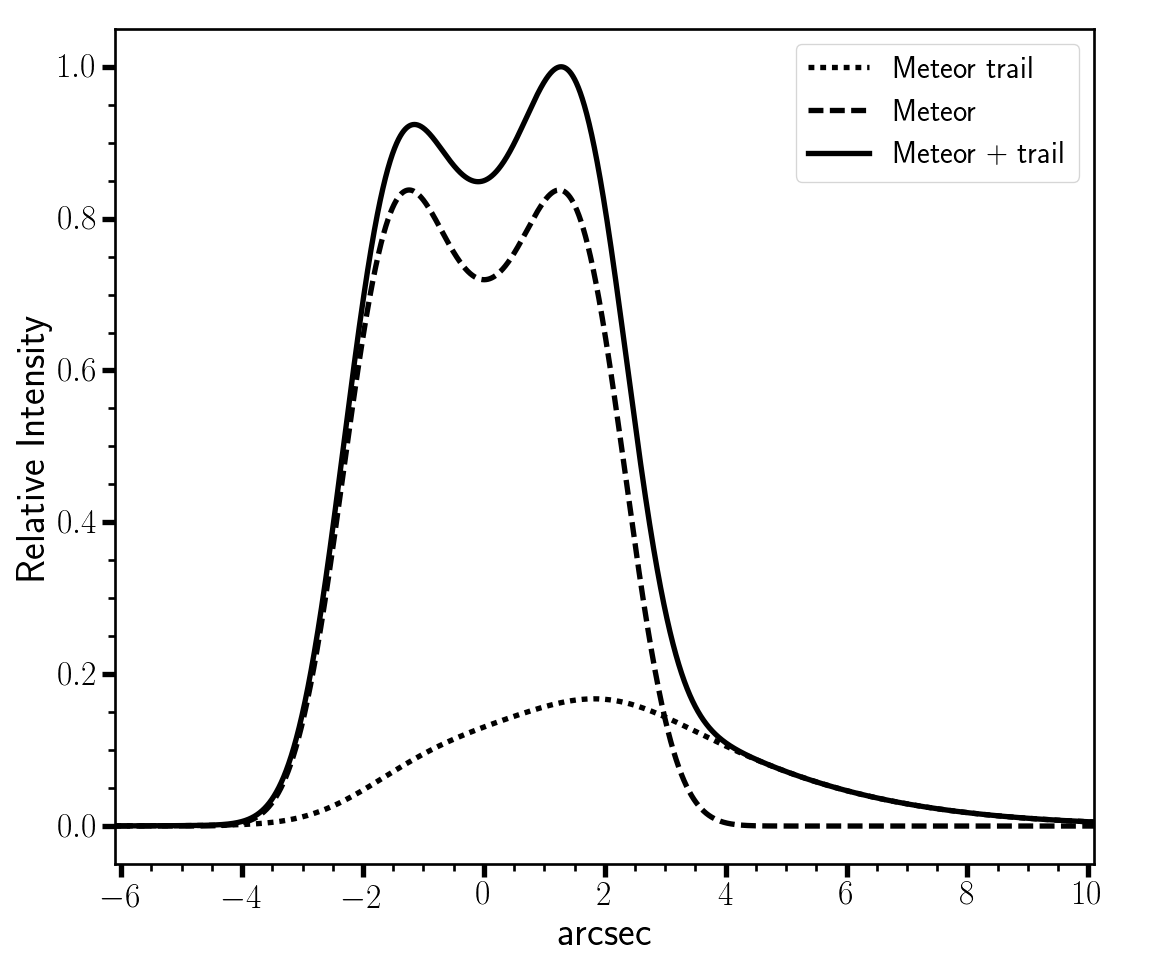}
  \caption{An example of the observed meteor track (solid line) as it would appear in an image from the SDSS telescope obtained as a sum of two contributions at 100~km distance: from a defocused meteor (dashed line) with 80\% of the peak brightness and from a defocused meteor trail (dotted line; see Fig. \ref{fig:meteor_traildriftSDSS}) with 20\% of the peak brightness. This example illustrate how the meteor trail deforms the pure meteor head brightness profile by deforming dominantly one side of the defocused two-peak meteor head profile. }
  \label{fig:addedMeteorTrailSDSS}
\end{figure}

\begin{figure}
  \centering
    \includegraphics[width=0.45\textwidth]{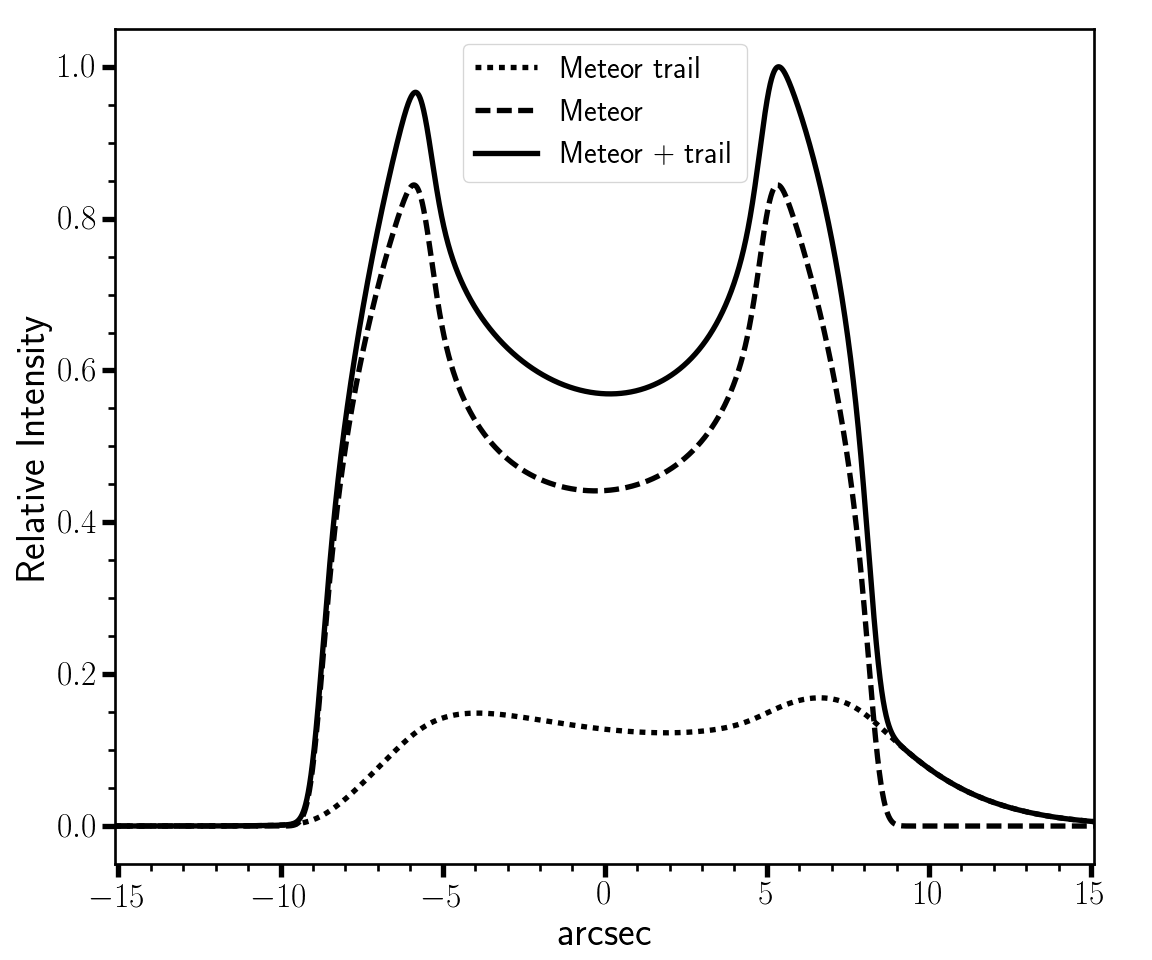}
  \caption{The same as in Fig. \ref{fig:addedMeteorTrailSDSS}, but for the LSST telescope using the meteor trail profile from Fig. \ref{fig:meteor_traildriftLSST}. In this case the trail's main disruption to the meteor head brightness is in reducing the depth of the central brightness dip, while the profile asymmetry is not very prominent. }
  \label{fig:addedMeteorTrailLSST}
\end{figure}

\section{Defocussing of meteor trails}
\label{sec:trail}

So far we have been discussing only the light produced by the meteor head. This is the main meteor component that illuminates the detector pixels as it flies through the field of view. However, meteors are also known by their glowing trail that they leave behind in the atmosphere. Although all meteors have a trail, made of the hot plasma left behind the meteor head, they are typically of much lower brightness than the head and disperse very quickly. The problem arises when the trail is bright and stays on the sky long enough to expose pixels to brightness levels comparable to the meteor head signal. This requires meteor trail modeling, which is a highly challenging topic due to diverse trail properties and their complex behavior in the atmosphere.

Such an ion trail model would need an initial non-Gaussian density profile \citep{Jones} that expands anisotropically as it depends on the geomagnetic field direction \citep{Oppenheim}. A detailed modeling of this process is out of scope of our paper, hence, we use just an illustrative case with a Gaussian intensity profile and the initial width of 0.8~m \citep[typically it is between $\sim$0.5~m and 3~m, depending on the altitude;][]{Stokan}. The model trail grows to the size of 10~m, when its surface brightness becomes too low for further observational consideration. The trail duration can vary significantly and it depends on many factors: the meteor velocity, the trail density, the ionosphere electron density, the presence of background winds and electric fields resulting from the ionospheric electrojets \citep{Dyrud}, etc. The basic assumption is that trails diffuse into disappearance exponentially in time as $e^{-t/\tau}$, with $\tau$ less than a second or just a few seconds \citep{Hocking}, although sometimes they can remain glowing for minutes \citep{Dyrud2007}. In our model we use $\tau$=1~s, with a constant total emitted light (integrated brightness profile).

The final result of our model is a brightness profile obtained by time integration of the evolving meteor trail profiles. In Figs. \ref{fig:meteor_trailSDSS} and \ref{fig:meteor_trailLSST} we show how the trail brightness changes in time and the final image brightness profile it produces. Initially the trail is very narrow and defocusing produces a central dip, but as the trail width grows, the dip disappears. The final integrated result is a very wide flat-top profile in SDSS and even wider profile in LSST, although with a central shallow dip. This profile needs to be added to the meteor head brightness profiles described in previous sections.

Another important feature of meter trail evolution is fast drift in position due to background winds that carry it away from the initial trajectory. To illustrate this effect and demonstrate its impact on the imaged meteor track, we introduce a drift speed to the meteor trail profiles in each simulation step. Unlike in the static case, here the final brightness profiles display prominent asymmetries (Figs. \ref{fig:meteor_traildriftSDSS} and \ref{fig:meteor_traildriftLSST}). When combined with the meteor head exposure, this drifting trail causes asymmetrical heights of the defocusing brightness peaks and extended brightness on one side of the meteor track profile. These effects are more prominent in SDSS images than in LSST (see Figs. \ref{fig:addedMeteorTrailSDSS} and \ref{fig:addedMeteorTrailLSST}). An example of SDSS meteor with such a brightness behavior is shown below in Fig. \ref{fig:frame-r-005973-5-0069}.

\begin{figure}
  \centering
    \includegraphics[width=0.46\textwidth]{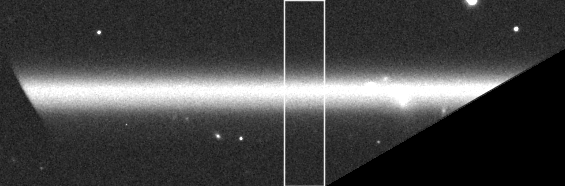}
    \includegraphics[width=0.46\textwidth]{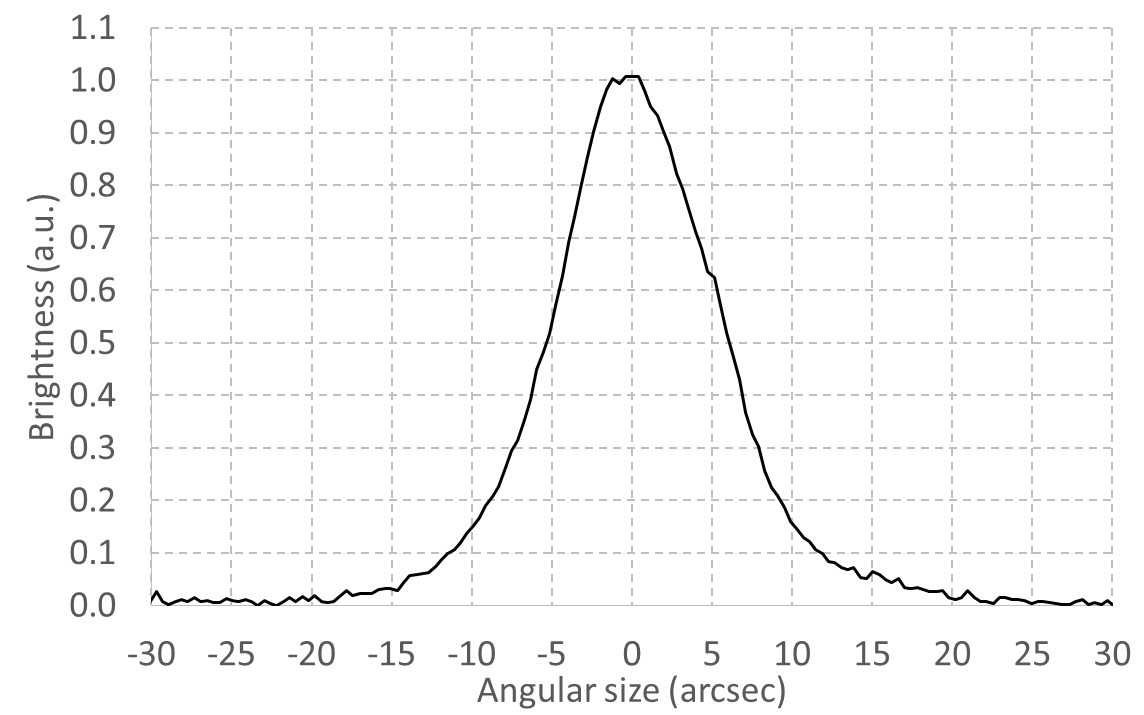}
  \caption{A meteor track in the SDSS image frame {\it frame-i-002728-3-0430} in the SDSS {\it i} filter. The upper panel is the meteor image extracted from the SDSS frame, rotated to horizontal direction and adjusted in brightness to appear visually enhanced. The lower panel is the median of pixel values in horizontal direction between the two vertical lines in the panel above. The unit of brightness can be arbitrary, although in this case the SDSS image frames are calibrated in nanomaggies (\url{http://www.sdss.org/dr12/help/glossary/\#nanomaggie}) per pixel and have had a sky-subtraction applied.}
  \label{fig:frame-i-002728-3-0430}
\end{figure}

\begin{figure}
  \centering
    \includegraphics[width=0.46\textwidth]{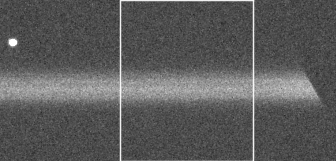}
    \includegraphics[width=0.46\textwidth]{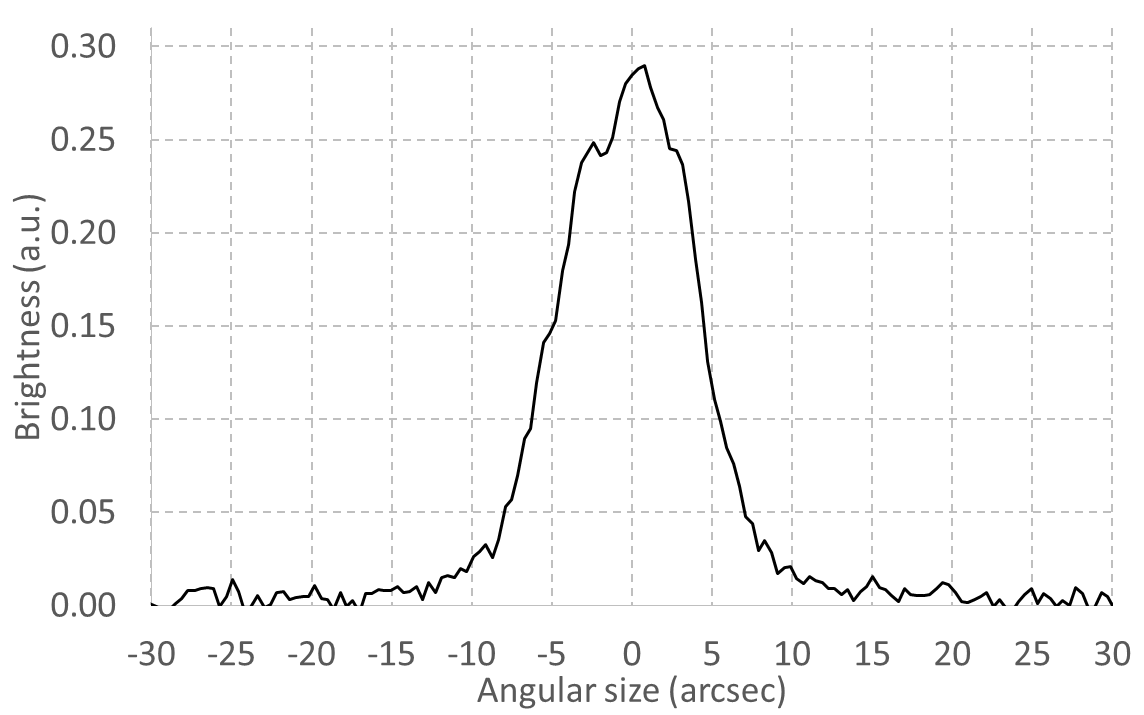}
  \caption{The same meteor as in Fig. \ref{fig:frame-i-002728-3-0430}, but here it passed over a CCD chip covered with the {\it u } filter. The frame is {\it frame-u-002728-3-0429}. Figure details are the same as in Fig. \ref{fig:frame-i-002728-3-0430}. The meteor here appears narrower than in the {\it i} filter and its profile is not smooth any more. }
  \label{fig:frame-u-002728-3-0429}
\end{figure}	

\begin{figure}
  \centering
    \includegraphics[width=0.46\textwidth]{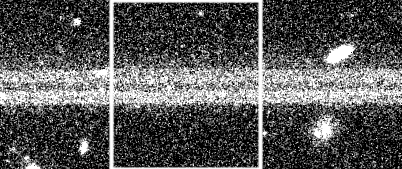}
    \includegraphics[width=0.46\textwidth]{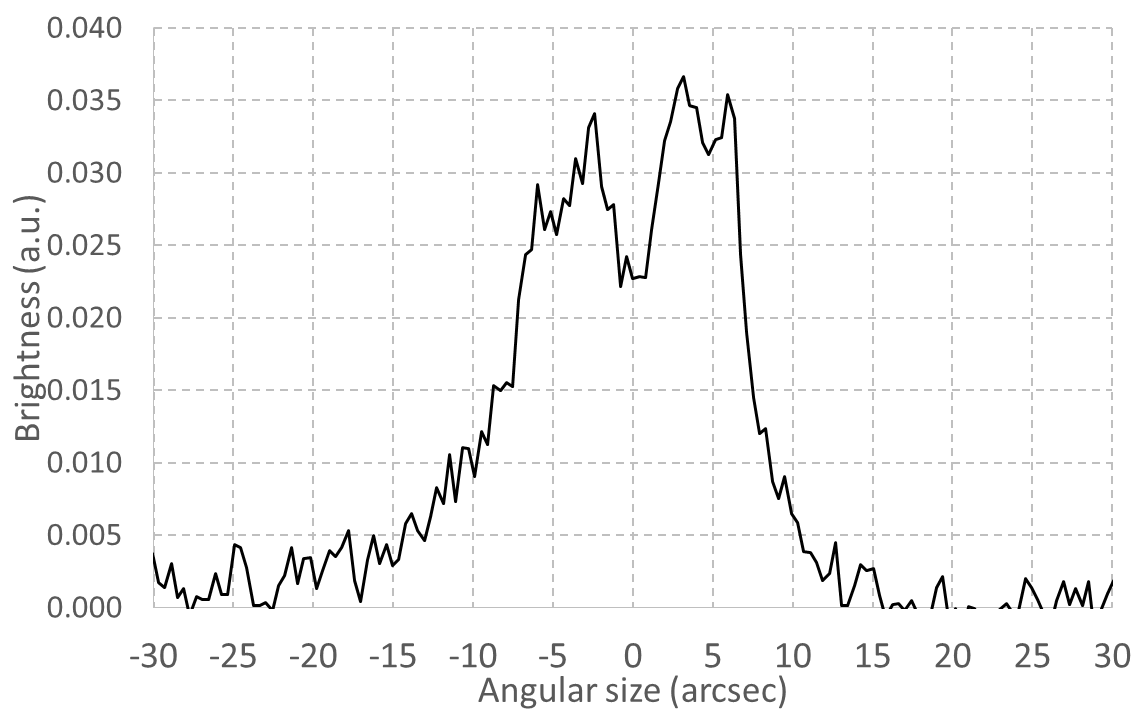}
  \caption{The same meteor as in Fig. \ref{fig:frame-i-002728-3-0430}, but here it passed over a CCD chip covered with the {\it g} filter. The frame is {\it frame-g-002728-2-0424}. Figure details are the same as in Fig. \ref{fig:frame-i-002728-3-0430}. When the meteor reached this filter it already produced two pieces that are seen here in the brightness profile as two broad peaks that have their own two smaller peaks (degraded by noise) produced by defocusing.}
  \label{fig:frame-g-002728-2-0424}
\end{figure}

\section{Defocusing of artificial satellites and space debris}
\label{sec:satellites}

One frequent source of long linear features in astronomical survey images are also satellites. They can be confused for meteors, thus, we need a method to differentiate between them. Satellites can be typically recognized by their smooth periodic light curve, but we need a more quantitative and reliable measure. It turns out that satellites leave tracks that have a significantly smaller FWHM than meteors.

Fig. \ref{fig:approximate} shows that the concern are satellites in the low Earth orbit (LEO). LEO altitudes range from 200~km to 2000~km, while we are most interested in the very low orbits up to $\sim$600~km.
The distribution of satellites and orbital decay times \footnote{\url{https://www.planet.com/pulse/keeping-space-clean-responsible-satellite-fleet-operations/}} show that only a small fraction of them are located at these altitudes as they can not maintain their orbit for a very long time. These low orbit altitudes are therefore populated with smaller satellites, especially nanosatellites. For example, \textit{Planet} company has a constellation of about 150 nanosatellites smaller than 0.5~m at altitudes below 700~km that will decay from their orbits in 25 years or less. Some of the brightest satellite flashes on the sky are produced by Iridium satellites that are 5~m in size (with solar panes) and at 780~km altitude, which still makes their trail FWHM significantly smaller than meteors in SDSS and LSST (Fig. \ref{fig:approximate}).

Another type of objects in these orbits are orbital debris, but they are also small, with the largest pieces smaller than $\sim$4~m  \citep{risk}. The only extreme is the ISS at altitudes between 330~km and 435~km that has about 100~m in size due to its large solar panels. It will be resolved and defocused, but this is the only such object in the sky and it has a predicted time of passage above the telescope.

The concept of satellites displaying smaller track widths than meteors was shown by \cite{Subaru} (see their Fig. 7) in images from the Subaru telescope. \cite{BiDS} also demonstrated this with SDSS images, where they used the angular speed as an additional independent measure (possible because of the drift scan mode of SDSS telescope operation). Their sample of SDSS tracks showed that satellites have angular speeds below $\sim 1^\circ/s$. While some meteors can appear very slow when projected on the sky if they move almost directly toward the camera, they are still faster than satellites (see their Fig. 3). At the same time, tracks from these satellites have their defocused angular size smaller than $\sim 3\arcsec$. In LSST this threshold size will be $\sim 6\arcsec$. Since the drift scan mode is typically not used by survey telescopes, the angular velocity cannot be determined and the classification by trail width is the key method for separating meteors and satellites.

\begin{figure*}
  \centering
    \includegraphics[width=\textwidth]{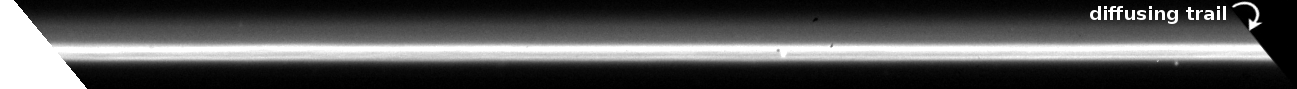}
    \includegraphics[width=\textwidth]{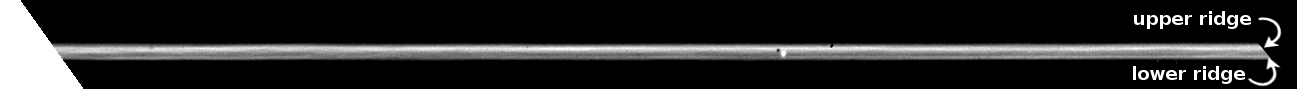}
    \includegraphics[width=\textwidth]{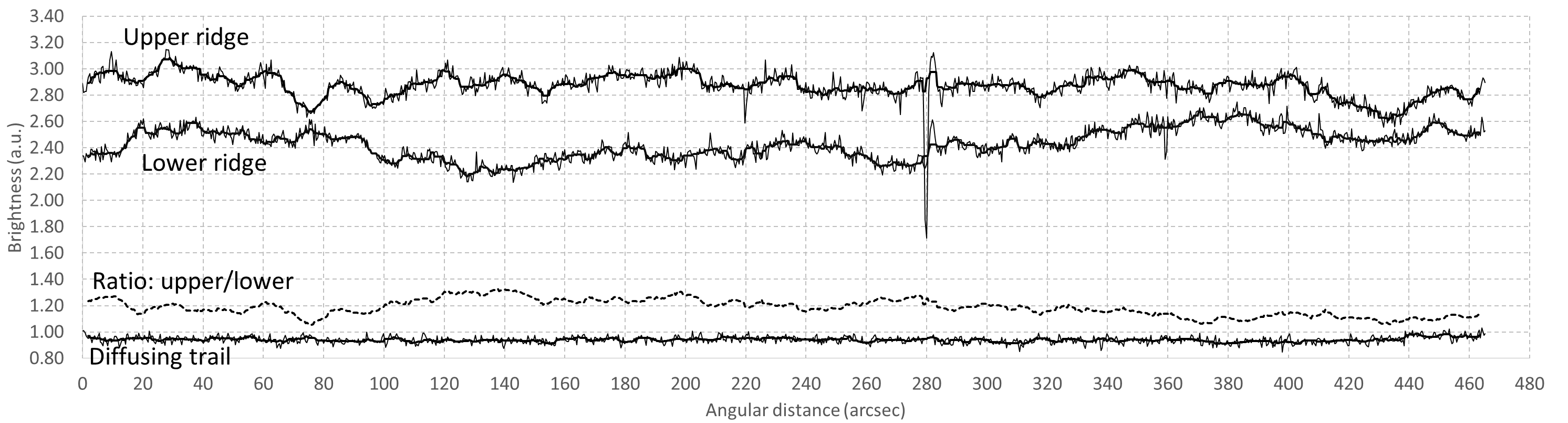}
  \caption{A meteor track in a SDSS image frame {\it frame-r-005973-5-0069} in the {\it r} filter. The upper panel is the meteor image extracted from the SDSS frame, rotated to horizontal direction and shown with adjusted brightness versions to appear visually enhanced. The first track (upper) is emphasizing the diffusing trail that was left behind the meteor. The second track (lower) is emphasizing two ridges produced by the defocus of meteor head. The lower panel shows brightness profiles (thin solid lines) following the meteor path along three lines: the two ridges of the meteor head and along the glow of diffusing trail (as marked in the upper panel). Thick solid lines are smoothed versions of the brightness profiles using 1D median filter of 4.36$\arcsec$ in width (11 pixels). The dashed line is a ratio of smoothed upper and lower ridge profiles. An example of brightness profile across this meteor track is given in Fig. \ref{fig:frame-r-005973-5-0069}.}
  \label{fig:LINES-frame-r-005973-5-0069}
\end{figure*}

\section{Examples from SDSS}
\label{sec:examples}

The SDSS camera consists of an imaging array of thirty 2048$\times$2048 photometric CCDs arranged into a grid of 5 rows and 6 columns \citep{Gunn}. The telescope scans the sky in drift scan mode in  five different filters (\textit{ugriz} photometric system), with each filter covering one row of CCDs. Filters wavelengths range from ultraviolet to near infrared (effective wavelengths are 3551~{\AA} for \textit{u}, 4686~{\AA} for \textit{g}, 6165~{\AA} for \textit{r}, 7481~{\AA} for \textit{i}, and 8931~{\AA} for \textit{z}) and their widths range from $\sim$600~{\AA} for \textit{u} to $\sim$1500~{\AA} for \textit{i}.

When a meteor pases through the field of view of SDSS camera, it exposes multiple CCDs and most often multiple filters (i.e. rows of CCDs). The diagonal of the CCD array is 2.7$^\circ$ long and the filters are ordered as \textit{riuzg} (from the leading to the trailing edge of the camera). Meteors appear quite different between filters depending on the meteor spectrum dominated by atomic and molecular lines \citep{Spectra}. This means that the meteor size can vary not only because it undergoes physical changes during the flight, but also because different emitting plasma components can be spatially distributed in different ways.

Figs. \ref{fig:frame-i-002728-3-0430}, \ref{fig:frame-u-002728-3-0429} and \ref{fig:frame-g-002728-2-0424} show snippets of a meteor that was detected by the SDSS camera on November 18, 2001, at 04:57:21.39 TAI. Its path indicates that it probably belongs to the Northern Taurid meteor stream. The altitude of the telescope pointing at that moment was 50.74$^\circ$ (this gives $\sim$130 km distance to the 100 km altitude point in the atmosphere) and the azimuth was 321.55$^\circ$. The meteor passed through \textit{riug} filters. If this was a Northern Taurid then the angle between the meteor path and the line of sight is $\sim$65$^\circ$. The appearance of meteor track differs significantly between the filters, which is an interesting topic to explore. In Fig. \ref{fig:frame-i-002728-3-0430} we show its brightness profile in the \textit{i} filter. The profile is similar to typical uniform disk models in Fig. \ref{fig:Disk_SDSS_objects} for a disk of $\sim$10~m in size if the meteor altitude was $\sim$100~km. This is an unusually large plasma ball for a meteor, but the filters that follow reveal what happened.

After exiting the \textit{i} filter, the meteor entered the \textit{u} filter. Between each row of CCD chips there is a $\sim$5$\arcmin$ large gap that corresponds to $\sim$200~m of meteor flight at 100~km altitude. The brightness profile of the meteor in the \textit{u} filter is shown in Fig. \ref{fig:frame-u-002728-3-0429}. The trail had become dimmer, slightly narrower and it is not smooth any more, but it shows some "grooves" in the track (seen as variations in the brightness profile). The meteor obviously shows some internal structure, which is a possible indication of meteor fragmentation. Indeed, by the time it had flown through the \textit{g} filter (almost 2~km of meteor flight later) it appears as two tracks of small dim pieces (Fig. \ref{fig:frame-g-002728-2-0424}). Each segmented track is consistent with a defocused object smaller than $\sim$1~m although the brightness noise is quite large. The separation between fragments is $\sim$6~m at 130~km distance. This now explains the initially large size in the \textit{i} filter - two fragments were initially bright, producing a large glowing plasma ball that mimicked a single big meteor, but as the brightness and size of individual plasma balls had waned their true nature as separate fragments emerged. This means we can interpret brightness grooves in the \textit{u} filter as fragments separated $\sim$3~m. If we assume $\sim$30~km/s speed of Northern Taurids then the transverse speed of fragments was $\sim$45~m/s when they entered the \textit{g} filter. The physical mechanism responsible for such a high transverse speed of fragments has not been identified yet \citep{fragment}.

\begin{figure}
  \centering
    \includegraphics[width=0.46\textwidth]{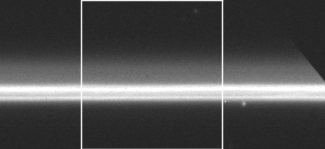}
    \includegraphics[width=0.46\textwidth]{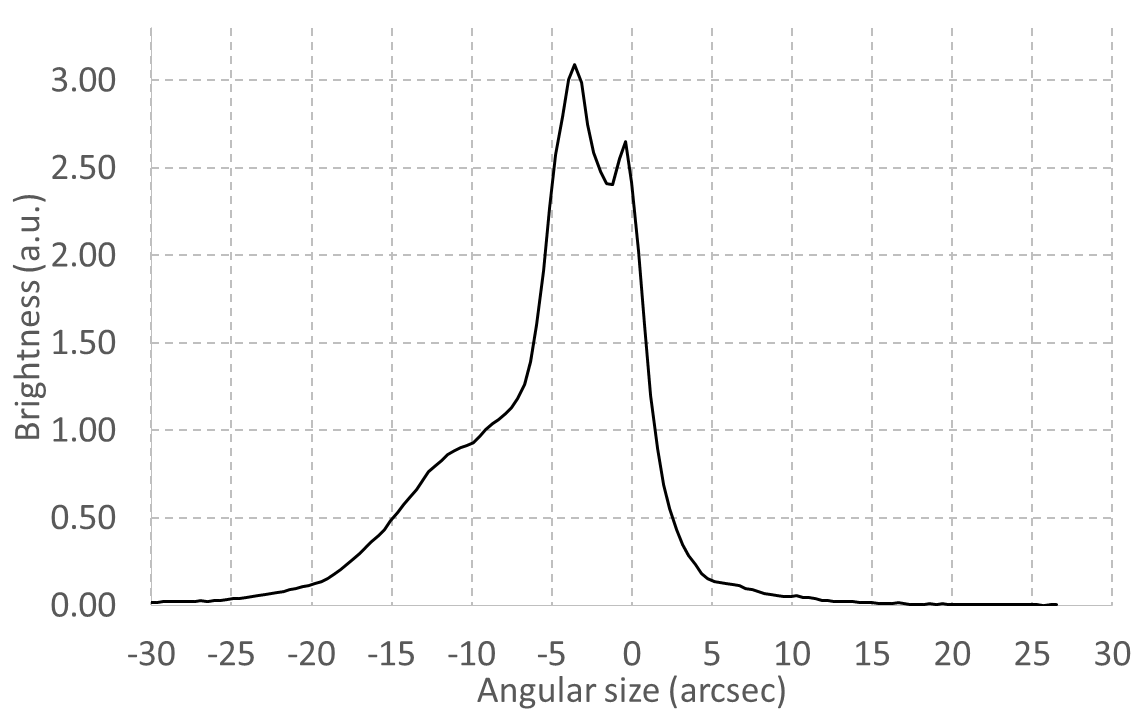}
  \caption{A fraction of the meteor shown in Fig. \ref{fig:LINES-frame-r-005973-5-0069}. The upper panel shows the meteor image with enhanced brightness, while the lower panel is the median of pixel values in horizontal direction between the two vertical lines in the panel above. Figure details are the same as in Fig. \ref{fig:frame-i-002728-3-0430}.}
  \label{fig:frame-r-005973-5-0069}
\end{figure}

The scientific motivation for studying such meteor images is the ability of big telescopes to resolve meteors and follow the evolution of meteor head structure as it flies through the atmosphere. Fig. \ref{fig:LINES-frame-r-005973-5-0069} is an example of such a meteor analysis. This meteor was imaged on January 6, 2006, and it entered the camera at 11:40:31.84 TAI. The telescope was pointing to 55.74$^\circ$ altitude and 1.45$^\circ$ azimuth. The path direction is not pointing to any major meteor stream, thus it is probably a sporadic. It left a track in \textit{riuz} filters. We show in Fig. \ref{fig:LINES-frame-r-005973-5-0069} the brightness variation along the track in the \textit{r} filter. The profile of a defocused meteor head is masked with the long lasting evolution of diffusing meter trail. The profile across the track (see Fig. \ref{fig:frame-r-005973-5-0069}) shows the width consistent with a meteor head of several meters in size, while on the other hand the existence of central brightness dip supports a size of $\sim$1~m or less. This is probably a confusion caused by trail diffusion, since it also shows two unequal peaks and a heavy brightness tail on one side that complicate the analysis. The peak brightness variations along the track in Fig. \ref{fig:LINES-frame-r-005973-5-0069} show complicated oscillations with the two peaks varying in different ways, while the profile along the diffusing trail is almost perfectly constant. This might be an indication that the wake immediately behind the meteor head is also highly variable in its initial fast expansion, but it turns into a smooth diffusion into the surrounding atmosphere at distances of $\sim$10~m or more away from the meteor path. This might be an indication of the over-dense meteor trail evolution that starts with an initially high temperature turbulent flow in the meteor wake on a millisecond time scale, which then evolves into a more dynamically stable ambipolar diffusion of the meteor trail into the surrounding atmosphere \citep{Silber}.

\begin{figure}
    \includegraphics[width=0.5\textwidth]{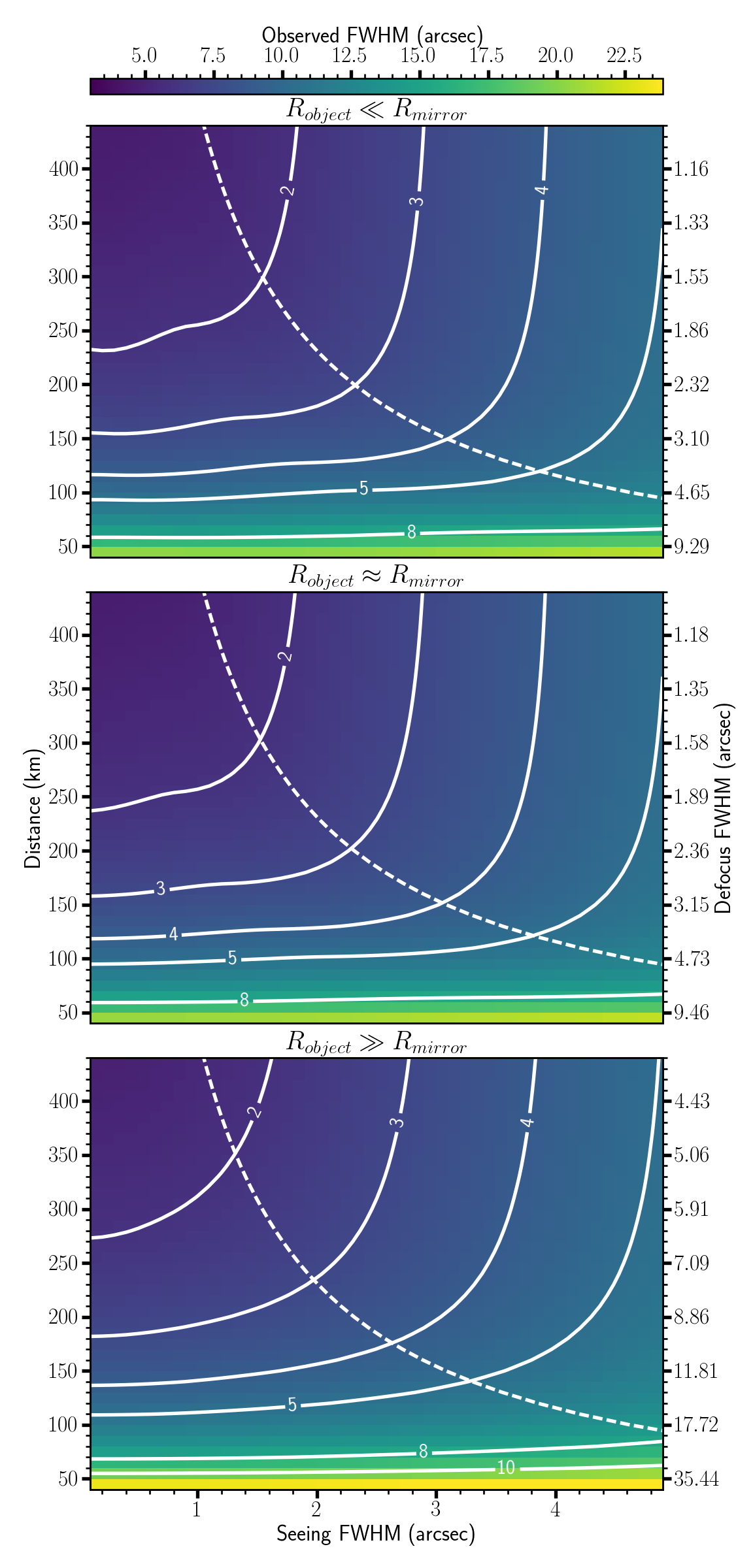}
  \caption{Plot of the observed FWHM (color scale and contours) as a function of distance and seeing for SDSS in three cases (from the top to bottom): point source, a uniform disk of $R_{meteor}$=0.9~m ($\approx R_{mirror}$) and a uniform disk of $R_{meteor}$=3~m ($\gg R_{mirror}$). The right axis shows the defocussing FWHM for distances indicated on the left axis. The dashed line represents FWHM for which the seeing is identical to the defocussing at a given height. Points above the dashed line are dominated by the seeing FWHM, while defocusing dominates points below the line.}
  \label{fig:ParameterSpaceSDSS}
\end{figure}

\begin{figure}
    \includegraphics[width=0.5\textwidth]{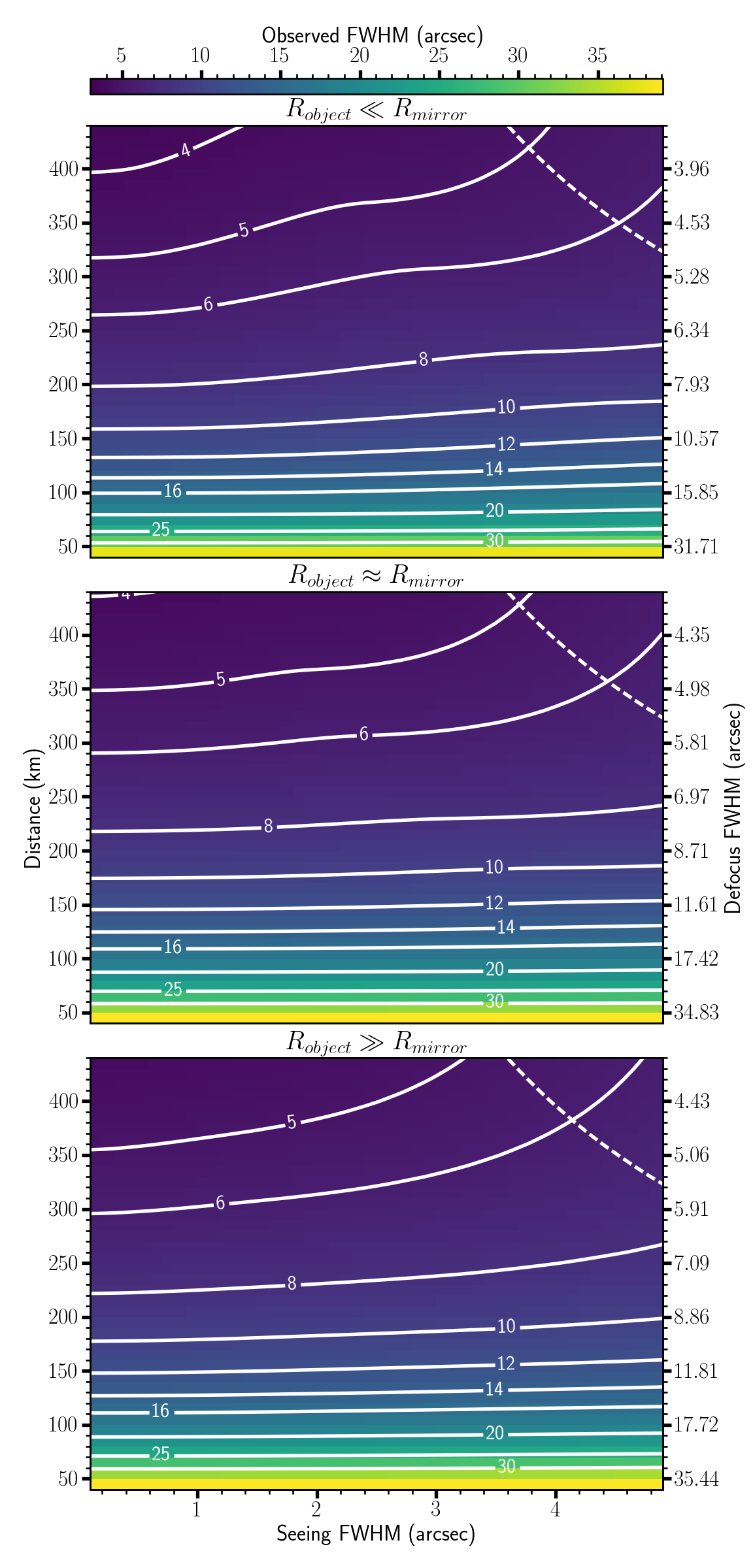}
  \caption{The same as in Fig. \ref{fig:ParameterSpaceSDSS}, but for LSST. The disk sizes are $R_{meteor}$=4~m  in the middle panel and  $R_{meteor}$=8~m in the bottom panel. The observed FWHM is almost completely dominated by the defocusing effect for the range of distances and seeing shown in these panels.}
  \label{fig:ParameterSpaceLSST}
\end{figure}

\begin{figure}
    \includegraphics[width=0.49\textwidth]{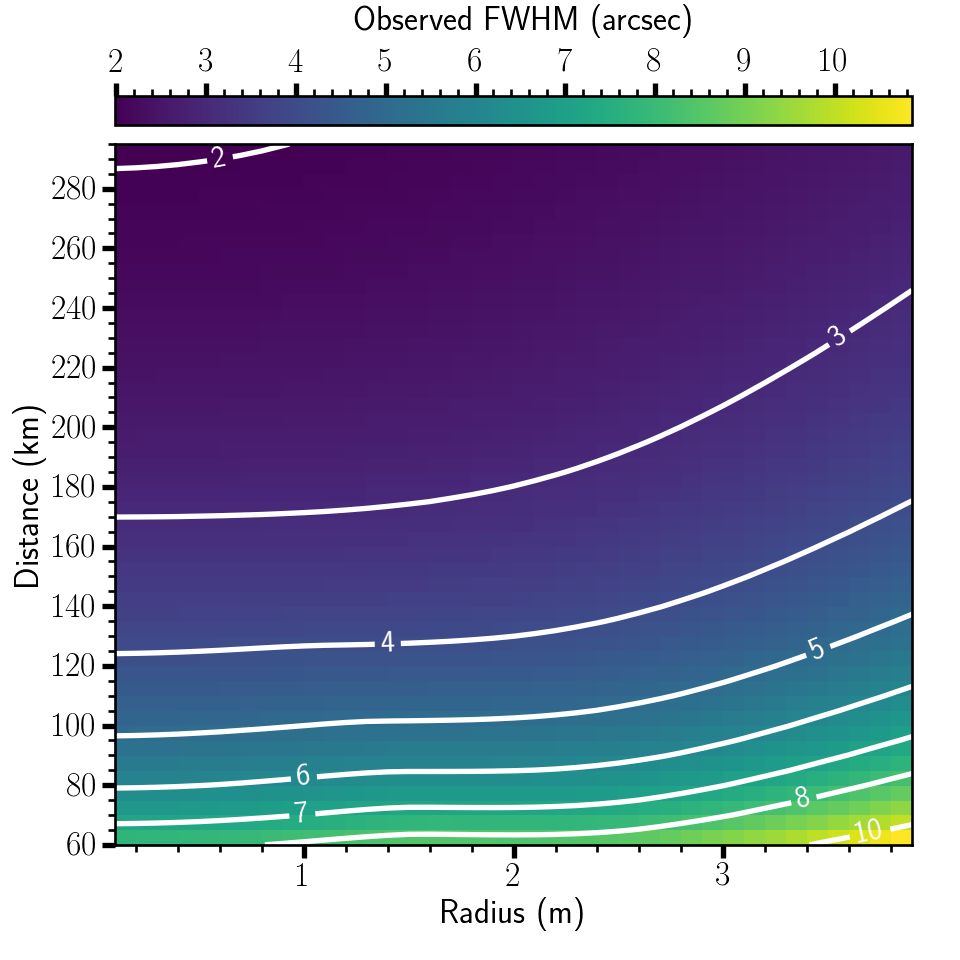}\\
    \includegraphics[width=0.49\textwidth]{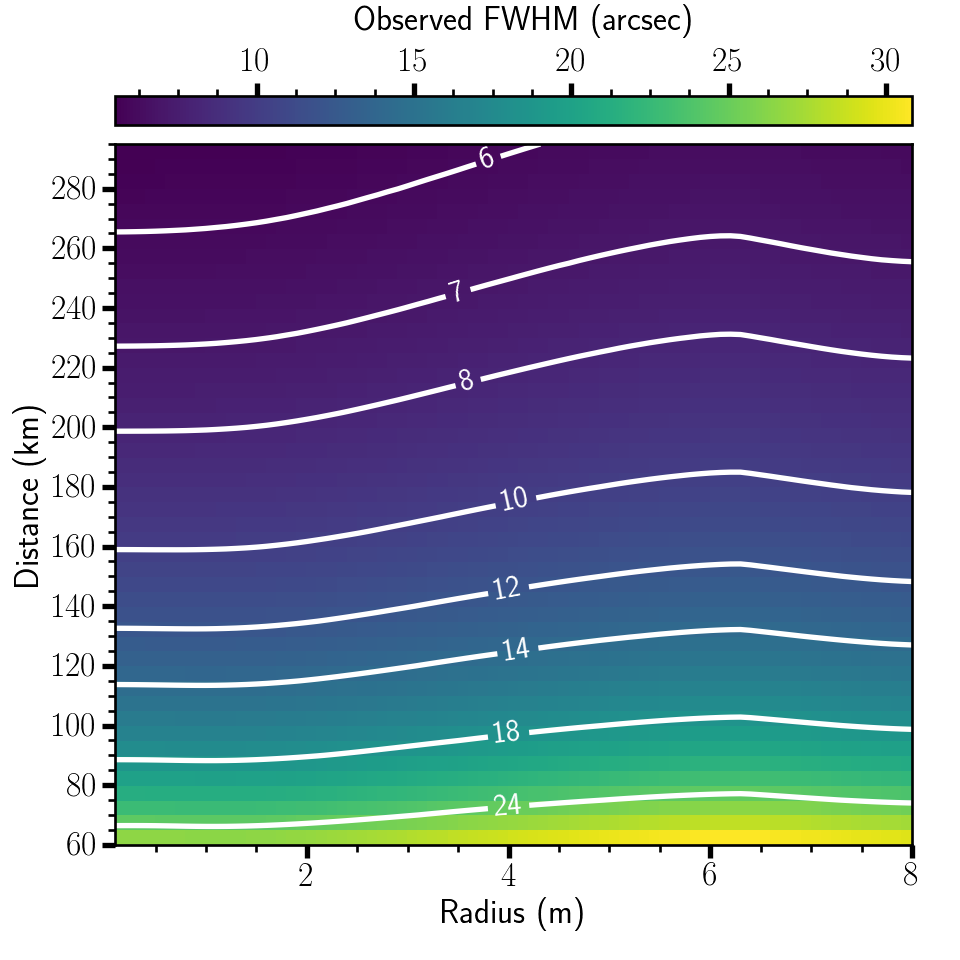}
  \caption{The observed FWHM (color scale and contours) as a function of a uniform brightness disk radius and meteor distance to the telescope. The top panel is for the case of SDSS (the seeing FWHM fixed to 1.48$\arcsec$) and the bottom is for LSST (seeing is 0.67$\arcsec$).}
  \label{fig:ObservedFWHMSRadii}
\end{figure}

\begin{figure}
    \includegraphics[width=0.48\textwidth]{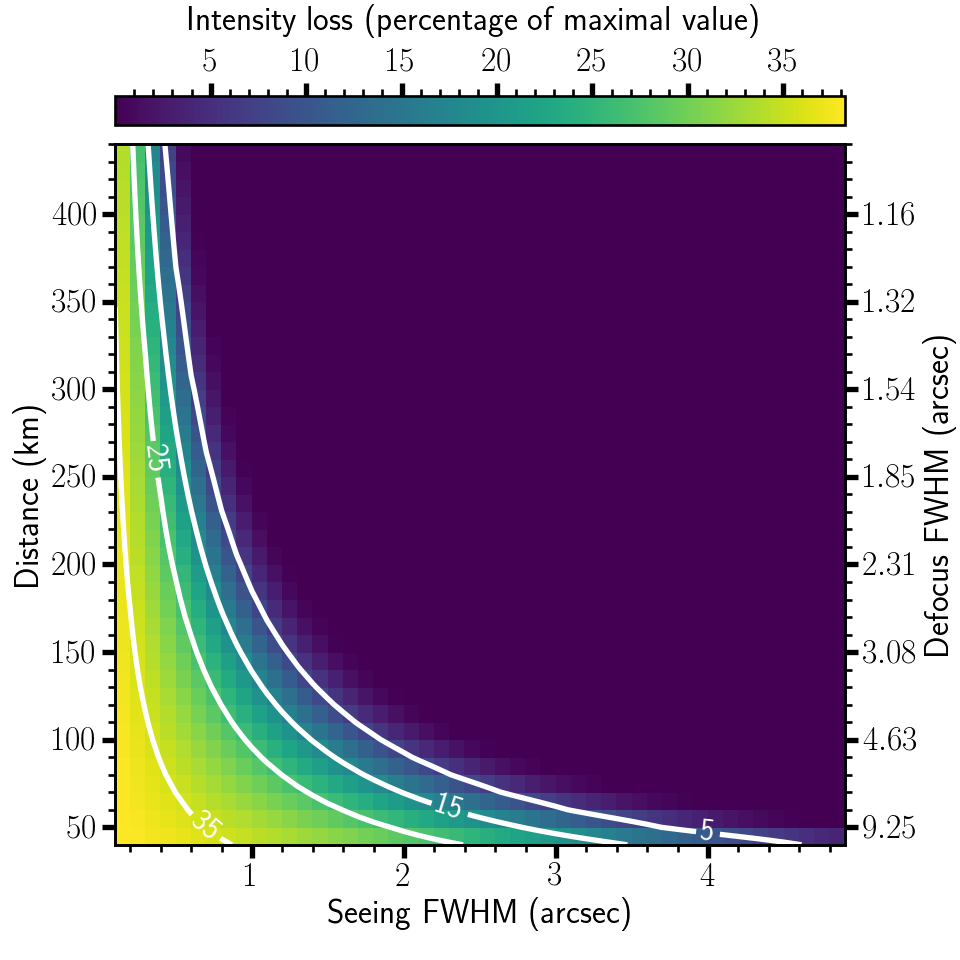}\\
    \includegraphics[width=0.48\textwidth]{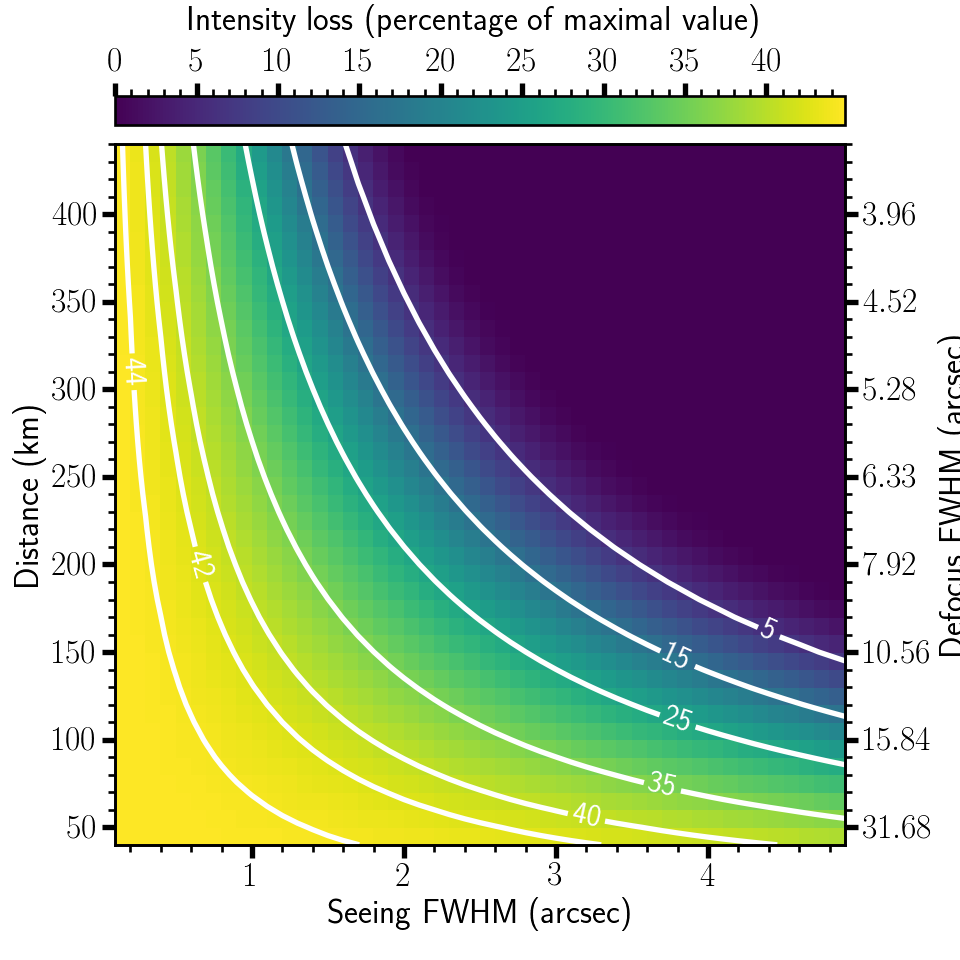}
  \caption{The strength of the central dip for a point source in the observed image profile measured as the intensity loss (color scale and contours) relative to the maximum brightness value in the profile (see e.g. Figs.\ref{fig:Point_seeing} and \ref{fig:Point_distance}). The panels show how the intensity loss depends on seeing and distance from the meteor in SDSS (top panel) and LSST (bottom panel). The right axis shows the defocusing FWHM for distances indicated on the left axis. }
  \label{fig:ParameterSpaceDEPTH}
\end{figure}

\begin{figure}
    \includegraphics[width=0.49\textwidth]{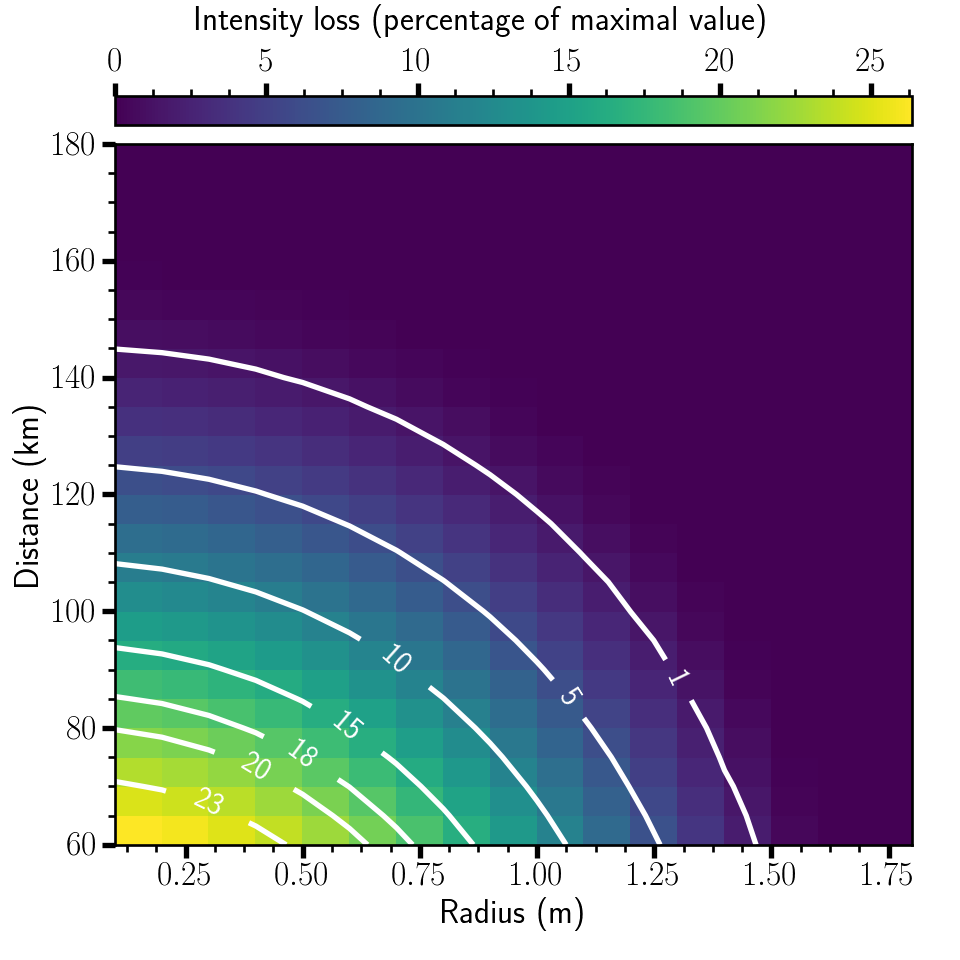}\\
    \includegraphics[width=0.49\textwidth]{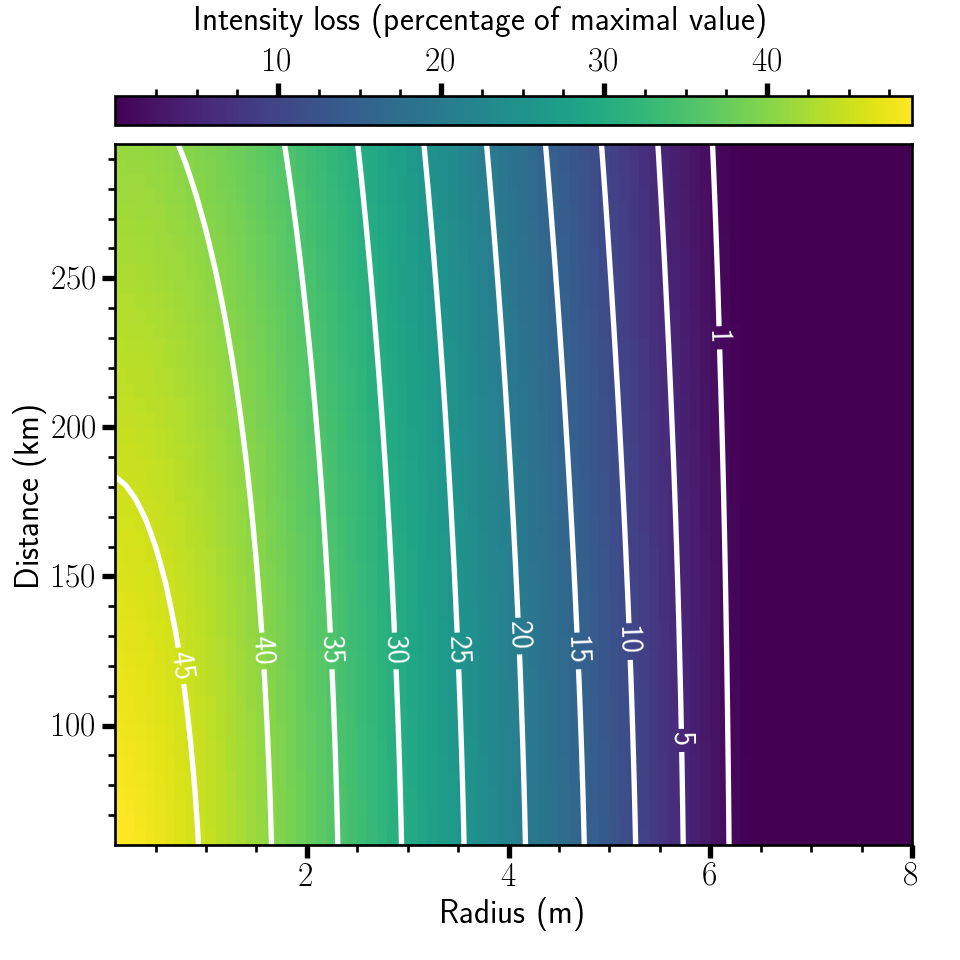}
  \caption{The same as in Fig. \ref{fig:ParameterSpaceDEPTH}, but here the horizontal axis shows the meteor head radius in a meteor model of a disk emitting uniformly from its surface. The seeing is set to 1.48$\arcsec$ for SDSS (top panel) and 0.67$\arcsec$ for LSST (bottom panel).}
  \label{fig:radiiDepths}
\end{figure}

\section{Breaking degeneracy between meteor distance and size}
\label{sec:discussion}

In the models of observed brightness shapes presented in Section \ref{sec:models} we can notice that seeing and defocusing contribute differently to the meteor image deformation. The main consequence of increased seeing is reduction of the depth of central dip in the observed profile, while defocusing most strongly impacts the observed FWHM. Thereby, according to the defocusing equation \ref{eq:pointDefocus}, we actually must consider how telescope aperture and distance to the meteor impact the observed meter image. This analysis is further complicated by the role of meteor head size, which is a model dependent variable because of our current limited knowledge of the visual shape of meteor heads. This forces us to use simplified model descriptions such as a uniform brightness disk or a Gaussian profile.

Overall we can describe the observed meteor head image profile with two quantities:
\begin{easylist}
\ListProperties(Hide=4,Hang=true,Style*=-- )
& observed FWHM,
& intensity loss at the central brightness dip (defined as a percentage of the maximum value in the brightness profile),
\end{easylist}
\noindent
while the variables that dictate those quantities originate from (in the most general terms) the meteor physics (e.g. meteor size, photon flux, brightness distribution), atmospheric and sky conditions (e.g. seeing, background brightness) and telescope properties (e.g. distance to the meteor, inner and outer radius of the primary mirror, detector pixel size).

Although we limited ourselves only to the SDSS and LSST telescopes, exploration of this parameter space gives us a general sense of what information can be extracted from the observed quantities. In Figs. \ref{fig:ParameterSpaceSDSS} and \ref{fig:ParameterSpaceLSST} we show how the observed FWHM depends of the distance and seeing in three cases: point source, the meteor size similar to the telescope aperture, and when the meteor is bigger than the aperture. We see that for distances relevant to meteors and under good seeing conditions the observed FWHM is dictated by the distance. This means we could, in principle, determine the meteor distance just from measuring its observed FWHM for an assumed meteor head size. In case of SDSS small variations of observed FWHM result in large (from the meteor physics point of view) changes in distance, thus the method would yield large errors on distance, but in LSST this procedure is more robust and it should be a viable method for extracting meteor distance. Moreover, Fig. \ref{fig:ObservedFWHMSRadii} shows that at typical meteor distances the changes in the meteor head size in the range of few meters introduce no more than about 10~km uncertainty ($\sim$10\% relative error) in the meteor distance (even less on the meteor altitude as the telescope is pointing away from the zenith). While in SDSS images it is difficult to measure the observed FWHM with such a high precision, in LSST images this will be relatively easy. For example, a step from 100~km to 110~km distance will produce a change in observed FWHM of $\sim$0.8$\arcsec$ under the expected seeing conditions at the observational site.

This means we could reach almost a kilometer precision on distance (given that the meteor signal is not strongly degraded by noise) if we can independently constrain the meteor head size. Fortunately, the central intensity loss in the observed brightness profile is giving us this possibility. Fig. \ref{fig:ParameterSpaceDEPTH} shows how the intensity loss varies with distance and seeing for a point source. The figure illustrates how both parameters strongly influence the depth of brightness dip; hence, if we know the seeing then this can be an independent constraint on the distance. If these two distance estimates do not match then it might be due to the meteor size. Fig. \ref{fig:radiiDepths} illustrates an example of how the radius of a uniform brightness disk influences the central intensity loss for a given value of seeing. The dependence is mixed with distance, especially in the SDSS example, but we can safely assume that the size of meteor plasma head is $\lesssim 2$~m if:
\begin{easylist}
\ListProperties(Hide=4,Hang=true,Style*=-- )
& SDSS: we clearly see the central dip
& LSST: the central intensity loss is larger than $\sim$44\% of the maximum brightness value in the profile.
\end{easylist}
\noindent
For meteor heads $\gtrsim$2~m in size there will be no distinct central intensity loss in SDSS images of meteors, while in LSST this limit of non-existent central dip is at meteor sizes $\gtrsim$10~m. However, for even larger objects their observed FWHM size is dictated by their physical size (see Figs. \ref{fig:DS_3Cases} and \ref{fig:Disk_LSST_distance}).

From this analysis we see that LSST images of meteor tracks will reveal important information about the meteor size and distance, while in SDSS this might be possible to achieve in some special cases of meteors. This work on meteors can be further combined with measurements of their emitted flux and direction of flight (to reveal a possible connection with meteor streams). In cases when meteors fly with an angle strongly inclined toward the telescope, differences between distances at different points of the meteor track can be measured and reveal the meteor entry angle. Or in an opposite case, when we know the entry angle, such as in meteors belonging to meteor showers, we can constrain changes in distance along the path and look for possible changes in meteor size.

Unfortunately, this optimistic view on the extraction of meteor size and distance is hampered by the light emission from meteor wake and trail that meteor leaves behind. They expose the same image pixels as the meteor head and deform the observed meteor head brightness profile. Any attempt to follow the above procedure on distance and size extraction will have to take a careful look at the trail contribution to the meteor image. This leads to a potentially complicated model dependent decomposition of different brightness contributions. Moreover, it has been shown that meteors fragment even at altitudes above 100~km \citep{fragment}, thus we need to be careful in the interpretation of meteor tracks not to confuse defocusing structures in the brightness profile with meteor fragmentation.

\section{Conclusions}

Meteor science has been so far almost completely marginalized in big astronomical sky surveys, probably due to lack of observational experience with meteors on telescopes of larger (several meters) aperture. Before the age of large digital sky surveys, such telescopes were designed for a small field of view, where meteors are extremely rare events treated as nuisances. This is changing now as sky surveys increase the probability of detecting meteors, with high-resolution sensors capable of resolving the physical size of meteor plasma structures.

We demonstrate this concept by detecting meteors in the SDSS survey, but we also make predictions for the upcoming LSST survey. As a part of this effort we have analyzed the effects that defocusing has on the observed meteor images. Defocusing is caused by meteors appearing too close to the telescopes focused to infinity. We present an analytic formula for the defocusing kernel, confirmed by ray-tracing simulations, needed to reconstruct the shape of brightness profiles of the tracks produced by meteors. The formula depends on the inner and outer radius of the primary telescope mirror and the distance to the meteor.

The meteor image is also affected by the atmospheric seeing, thus we explore how the interplay between seeing and defocusing affects the observed meteor brightness profile. The most prominent feature of defocused meteors is a double-peaked brightness profile, which should not be misidentified as meteor fragmentation. The depth of the central brightness dip depends on several factors, from the radius of primary mirror to meteor distance and seeing. Another important parameter in the image reconstruction is the physical size of the glowing meteor plasma ball. We show how meteor size influences the observed defocused image, but also how the image depends on the model chosen to represent the glowing plasma distribution. Understanding of meteor defocusing allows identification of meteor fragmentation and brightness flickering (due to differential ablation and/or turbulent wake).

The modeling is further complicated by a superposition of three meteor components that emit light - the meteor head plasma surrounding the meteoroid, the wake immediately behind the head and the trail of ionization left behind that can glow for some time and drift due to atmospheric winds. We show some examples of the interplay between the defocused head brightness profile and the defocused trail, but this remains a problem that requires further analysis.

We argue that the size and distance of satellites and space debris is such that their defocused tracks are narrower than meteors, which enables separation between these two classes of objects. This has been shown already using SDSS data, but now we predict that in LSST this difference will be even more visible. Another interesting result is a method for independently estimating the size and distance of meteors imaged by LSST. The method exploits differences in contributions to the defocus effect - the distance impacts mainly the width of the meteor track, while the size affects the central dip in the image profile as long as the meteor is below some critical value required for creating the dip.

Our study shows that large aperture sky survey telescopes are highly valuable instruments for studying meteors, but image interpretations will be model dependent because of the interplay between the light contributions from meteor head, wake and ionized trail. Defocusing complicates the meteor observations, but some method of deconvolution or forward modeling might enable reconstruction of high resolution images of meteors.

\label{sec:conclusion}

\section*{Acknowledgments}
We thank the Technology Innovation Centre Me\dj{}imurje for computational resources.
DB, DV and \v{Z}V acknowledge being a part of the network supported by the COST Action TD1403 "Big Data Era in Sky and Earth Observation", which helped with discussions of the applicability of our research.

We would also like to thank Sloan Digital Sky Survey for the free and open data access and accompanying detailed documentations. Funding for SDSS-III has been provided by the Alfred P. Sloan Foundation, the Participating Institutions, the National Science Foundation, and the U.S. Department of Energy Office of Science. The SDSS-III web site is http://www.sdss3.org/.

SDSS-III is managed by the Astrophysical Research Consortium for the Participating Institutions of the SDSS-III Collaboration including the University of Arizona, the Brazilian Participation Group, Brookhaven National Laboratory, Carnegie Mellon University, University of Florida, the French Participation Group, the German Participation Group, Harvard University, the Instituto de Astrofisica de Canarias, the Michigan State/Notre Dame/JINA Participation Group, Johns Hopkins University, Lawrence Berkeley National Laboratory, Max Planck Institute for Astrophysics, Max Planck Institute for Extraterrestrial Physics, New Mexico State University, New York University, Ohio State University, Pennsylvania State University, University of Portsmouth, Princeton University, the Spanish Participation Group, University of Tokyo, University of Utah, Vanderbilt University, University of Virginia, University of Washington, and Yale University.
\nocite{*}


\appendix
\section{The fine structure in defocused surface brightness}
\label{AppendixA}

\begin{figure}
    \includegraphics[width=0.48\textwidth]{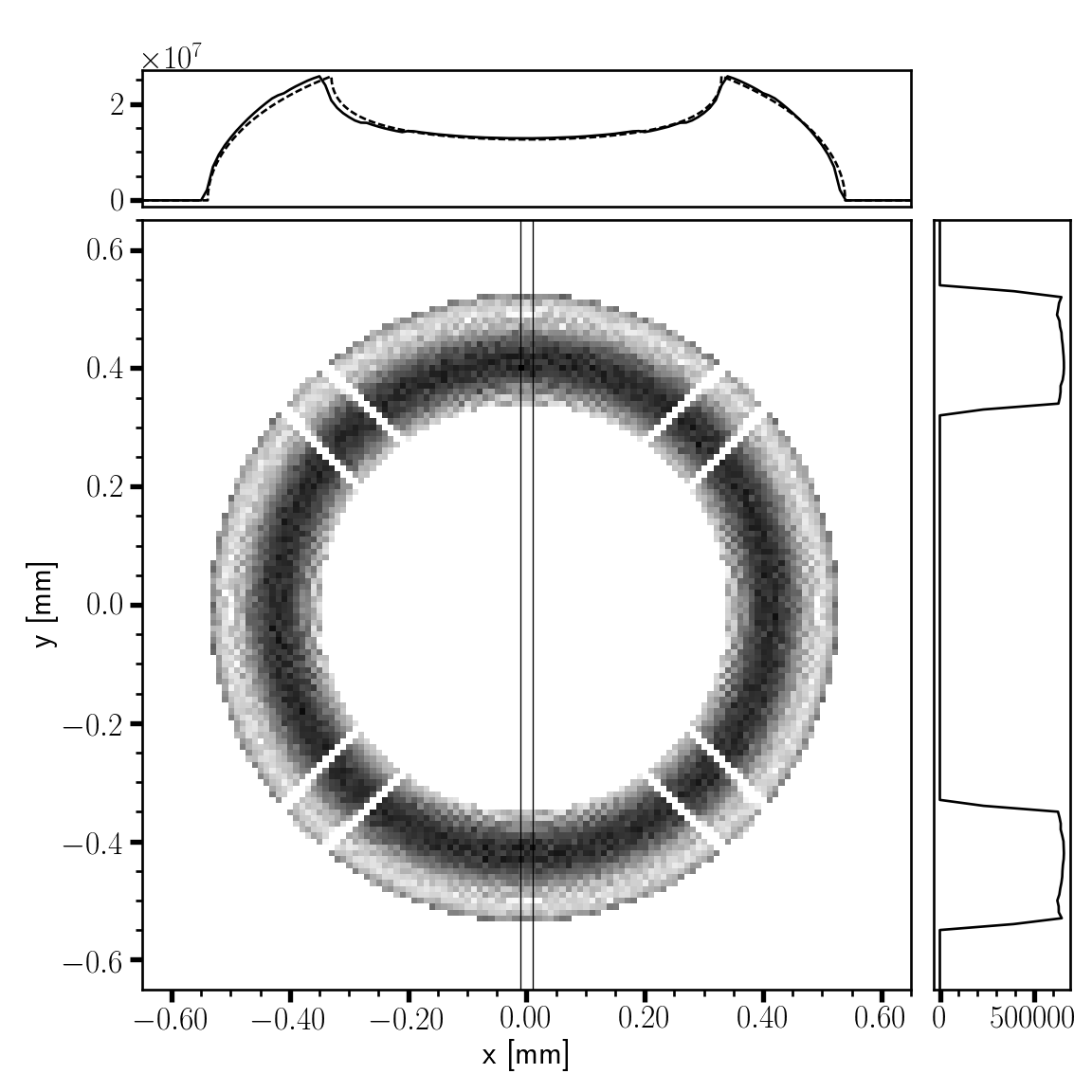}
  \caption{The central panel shows a simulated point source without seeing at 80~km distance to the telescope. The image contrast is adjusted to show pixels with count values 7.2 times larger than the count mean. The Spider's legs (visible as double lines at 45$^\circ$ angle) and mirror holders (small rectangles at the outer mirror edge horizontally and vertically) are clearly visible. The vertically integrated brightness profile is shown in the top panel (dashed line) and compared with the theoretical equation \ref{eq:pointDefocus} (solid line). The right panel shows values along the vertical cross section (marked by two thin lines in the central panel positioned at $0\pm10\mu m$).}
  \label{fig:fineStructure}
\end{figure}

The underlying assumption in equation \ref{eq:pointDefocus} is that every mirror section contributes equally to the final brightness distribution of an observed object. However, this is not necessarily true when second order effects are considered, as demonstrated in Figure \ref{fig:fineStructure}. The figure shows a simulated point source object with an apparent magnitude of 7.9, exposed for 15~s and placed at 80~km distance from the telescope. The object was simulated without seeing, vibrations or blurring of any kind, such that the end result corresponds purely to the geometric transformation of the aperture function. Typical pixel contains $325\pm\sqrt{325}ke^{-}$, consistent with the expected Poisson distributed shot noise. This is approximately 3 times the LSST pixel well depth.

In order to visually emphasize the distortions affecting the object's observed surface brightness, the image contrast is adjusted such that only pixels with the count value 7.2 times larger than the average pixel count are displayed. It is apparent that, apart from obstructions such as the telescope's spider, there exists a non-uniform radial distribution function. However, the variation between the maximum and minimum values along a vertical cross section is $\sim$5.5\% (the right panel in Figure \ref{fig:fineStructure}). This variation is small enough for the analytic equation \ref{eq:pointDefocus} to fit very closely the simulated profile (the top panel in Figure \ref{fig:fineStructure}), especially after applying the atmospheric seeing that blurs the image.

\section{Off axis image distortion effects.}
\label{AppendixB}

\begin{figure}
    \includegraphics[width=0.48\textwidth]{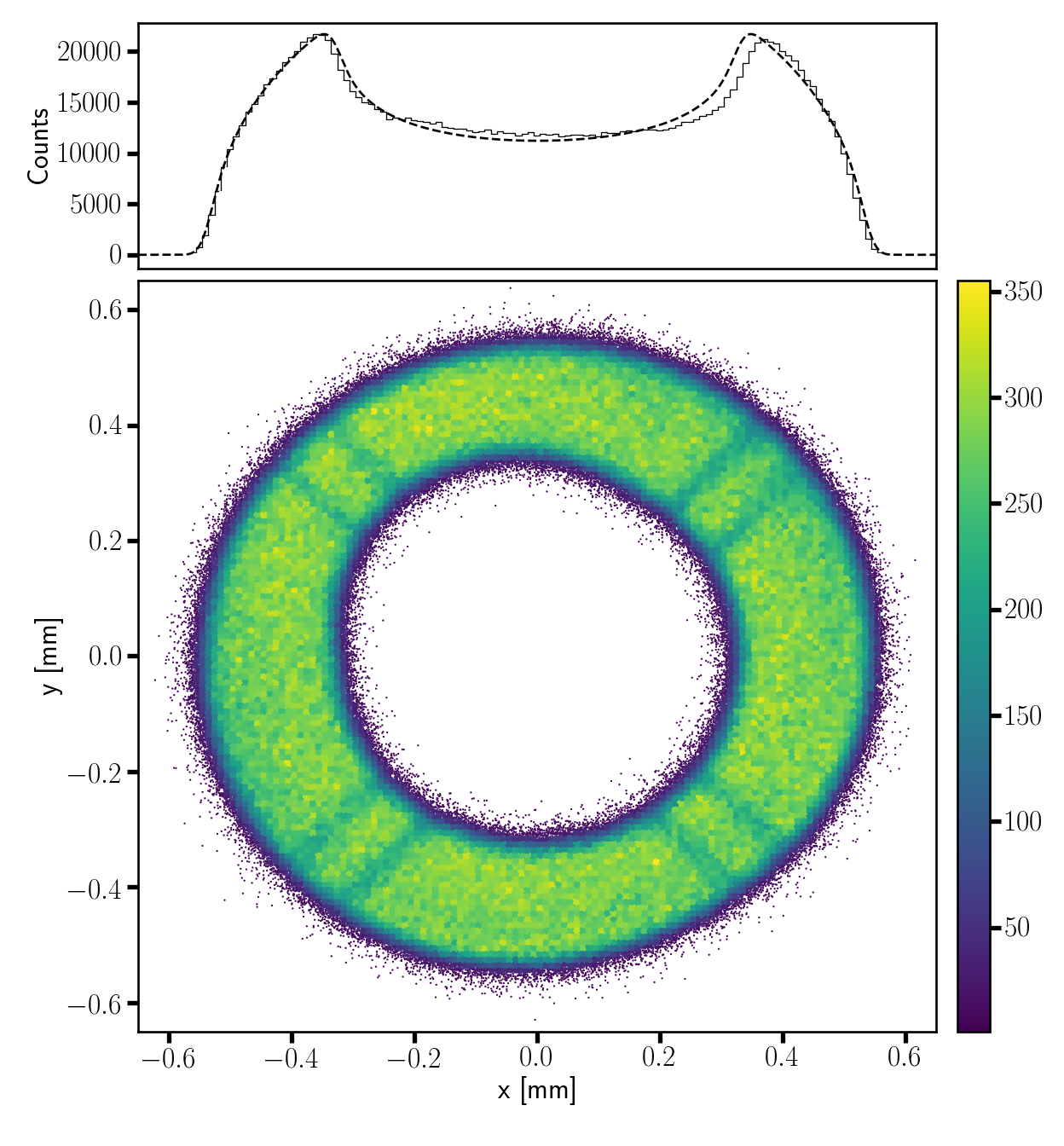}
  \caption{An off-axis simulated defocused point source of $\sim$14$^{th}$ AB magnitude at the distance of 80~km and 0.67$\arcsec$ seeing. Field distortions warp the otherwise circular symmetry of the defocusing effect (see Fig.\ref{fig:fineStructure}). The histogram (top panel) shows comparison between the simulated image integrated vertically (full line) and the theoretically predicted profile (dashed line).}
  \label{fig:offaxisdistort}
\end{figure}

Important additional effects that need to be considered in a high-precision modeling or measuring of imaged trails are the distortion effects that occur when objects are far from the optical axis of the telescope. We illustrate this with an example shown in Fig. \ref{fig:offaxisdistort}. This is a simulated 14$^{th}$ magnitude point source at 80~km distance, positioned approximately 1.5$^\circ$ off optical axis and under the seeing of 0.67$\arcsec$. The circular symmetry of the defocusing effect is now distorted into an egg-like pattern. These distortions will change as the meteor is observed through different parts of the atmosphere and different off-axis angles during its long flight on the sky.

\label{lastpage}

\end{document}